\documentclass[letterpaper,notitlepage,twocolumn]{revtex4-2}

\usepackage{%
	multirow,	
	hyperref,	
	units,		
	amsmath,
	amssymb,
	enumerate,	
	calc,		
	subcaption,	
	pgfplots,   
	pgfplotstable, 
	pgf-pie,    
	}

\hypersetup{colorlinks=true,linkcolor=blue}

\newcolumntype{C}[1]{>{\centering\arraybackslash}p{#1}}

\usetikzlibrary{backgrounds}
\usepgfplotslibrary{statistics}
\pgfplotsset{every boxplot/.style={solid,mark=x,}}

\definecolor{colorClarifying}{HTML}{ffb958}
\definecolor{colorExplaining}{HTML}{f54a45}
\definecolor{colorListening}{HTML}{8bdb91}
\definecolor{colorCDialogue}{HTML}{238538}
\definecolor{colorODialogue}{HTML}{145522}
\definecolor{colorObserving}{HTML}{00b0d4}
\definecolor{colorActiveObs}{HTML}{00b0d4}
\definecolor{colorPassiveObs}{HTML}{a9e7f5}
\definecolor{colorPresenting}{HTML}{0081d3}
\definecolor{colorIdeas}{HTML}{fd5cca}
\definecolor{colorNotInteracting}{HTML}{7e7e7e}
\pgfplotscreateplotcyclelist{SgWcColorList}{%
colorClarifying\\
colorExplaining\\
colorListening\\
colorCDialogue\\
colorODialogue\\
colorObserving\\
colorIdeas\\
colorNotInteracting\\
}

\definecolor{colorWC}{HTML}{800000}
\definecolor{colorInd}{HTML}{ffdab9}
\definecolor{colorSG}{HTML}{483d8b}
\pgfplotscreateplotcyclelist{GroupColorList}{%
colorWCg\\
colorInd\\
colorSG\\
colorNotInteracting\\
}

\begin{document}

\title{Examining teaching assistant pedagogies in traditional laboratories and recitations}
\author{Eric M.\ Hickok and Cassandra A.\ Paul}
\affiliation{Department of Physics \& Astronomy, San Jos\'{e} State University, One Washington Square, San Jose, CA 95192}

\begin{abstract}

Physics education research has consistently shown that students have higher learning outcomes when enrolled in active learning courses. However, while there is a lot of literature describing the difference between the two extremes of traditional vs.\ active learning courses, research has also shown that many classrooms actually lie on a spectrum rather than firmly falling into one category or the other. Understanding the pedagogical landscape is important for curricular development and dissemination, as well as targeted professional development efforts. Replicating and expanding on a study done by West \textit{et al.}~\cite{West2013Variation}, we observe graduate student Teaching Assistants (TAs) facilitating introductory physics labs and recitations using the Real-time Instructor Observing Tool (RIOT). We confirm West's finding of large variation between TAs' interactions during recitation sessions, but we also find that TAs facilitating traditional labs display fairly similar interaction profiles to each other. Additionally, we find that both the recitations and lab sessions we studied displayed very different interaction patterns from the CLASP ``Discussion/Labs'' studied by West \textit{et al.} Specifically, we find that the amount of time instructors spend observing students is a key distinguishing characteristic between the traditional settings and the CLASP curriculum. We discuss the pedagogical features of each of the different learning environments as captured by RIOT. We share our results as a snapshot of the interactive elements of an introductory physics course at a four-year, public, master's granting institution situated in a discussion on implications for reform efforts.
\end{abstract}
	
\maketitle

\section{Introduction}

Many introductory physics courses also have a labs and/or recitation component. These course elements are typically separate from the course meeting time and are often smaller than the lecture portion of the course. Labs and recitations are also often taught by graduate student Teaching Assistants (TAs). There is little argument from physicists that labs are an essential component of physics instruction. However, if you ask instructors what the purpose of labs are, you will get a variety of answers. For example, labs might be used to reinforce physics concepts, learn experimental skills, learn about uncertainty, and/or build collaboration and communication skills~\cite{holmes_introductory_2018}. Recitations (or workshops or discussion sections as they are also called) are another type of course supplement that serves a variety of purposes. They can be used to practice problem-solving learned material, learn new physics theories, or build physics community. 

Because labs and recitations sections are usually smaller than the overall course they serve, they can provide an excellent opportunity for instructors to interact with students one-on-one and in small groups. This also means that there is a greater opportunity for active learning activities and pedagogies to be employed. 

In this paper, we aim to understand the interactions that take place between the instructor and their students in traditional labs and recitation sections. We show that instructors in these course sections spend significantly less time interacting with their students indicating that they may not be taking full advantage of the opportunity to employ active learning techniques.

\subsection{Variation of Pedagogical Strategies in Active Learning}

A common finding from physics education research is that students tend to learn more when they are actively engaged in the learning process~\cite{hake, mazur, VonKorff2016, kramer_establishing_2023, Freeman2014b}. Meltzer and Thorton describe techniques that support active student learning as those which ``require all students to express their thinking through speaking, writing, or other actions that go beyond listening and the copying of notes, or execution of prescribed procedures''~\cite{meltzer_resource_2012}. While there are other shared qualities of active learning -- such as emphasizing conceptual understanding and utilizing multiple representations -- the idea that all students take on active participant roles in the classroom is central to active learning. Therefore, in order to support active student learning, instructors employ pedagogies that go beyond explaining physics content, and additionally create space for students to share their thinking and reasoning and get feedback on those ideas.

Pedagogy is a crucial element in the implementation of active learning, but instructors implement pedagogical techniques in different ways, often picking and choosing only parts to implement, sometimes to the detriment of the technique. For example, Turpen \textit{et al.} observed six faculty teaching six different introductory physics courses~\cite{Turpen2010Construction} using \emph{Peer Instruction}~\cite{mazur:PI} and found significant variation in how the professors interacted with their students during the administration of the clicker questions. Only half of the professors ``left the stage'' while students worked on and discussed the questions. There was a large amount of variation in how often the professors answered student-initiated questions and how they engaged in discussion with the students. They also found wide variation in the amount of time allotted for students to answer clicker questions and how the answers to the clicker questions were discussed. Some instructors frequently asked for student explanations while others did so rarely. Similarly, some instructors discussed incorrect answers occasionally while others did so rarely. Similar findings were obtained by West \textit{et al.}~\cite{West2013Variation} and later Wilcox \textit{et al.}~\cite{wilcox} who investigated the interactions between instructors and students engaging with the \emph{CLASP} curriculum~\cite{Potter:Sixteen} and the \emph{University of Maryland Open Source Tutorials}~\cite{MarylandTutorials,MarylandTutorials2}, respectively, using the Real-time Instructor Observing Tool (RIOT)~\cite{Paul2018b}. Both the West and Wilcox research teams found a wide range of instructor interactions with students. For example, West found that instructors spent anywhere from 0\% to more than 70\% of their whole class discussion time lecturing (explaining physics content). Since the purpose of the whole class discussions in CLASP are to have students discuss their ideas, the curriculum was not necessarily enacted as intended~\cite{West2013Variation}. These examples are corroborated by Dancy \textit{et al.} who state one of their key findings from their 2016 study on faculty implementation of Peer Instruction as ``Faculty generally modify specific instructional strategies and may modify out essential components'' of active learning curricula~\cite{Dancy2016}. Smith \textit{et al.} used the COPUS to examine pedagogies across a sample of STEM courses and also found large variability in pedagogy, implying that rather than falling into neat pedagogical boxes (such as traditional and active learning), pedagogical techniques exist as a spectrum at the classroom level~\cite{smith_campus-wide_2014}. Similarly, a recent study of over 1000 faculty surveys indicates that while active learning use has increased over the years, the extent to that it is used in the classroom varies significantly by instructor~\cite{dancy_physics_2024}.

The impact of the variation of pedagogical implementation in active learning courses is to some extent an open area of research~\cite{Docktor2014}. However, since research so clearly indicates that active learning environments are superior to traditional environments on many different measures~\cite{hake, mazur, VonKorff2016, kramer_establishing_2023, Freeman2014b}, it is to some extent predictable that differences in implementation will lead to differences in student outcomes.

\subsection{Examining Pedagogy: Labs and Recitations}

Recently there have been several studies examining the pedagogical landscapes of labs and recitations~\cite{wilcox, saitta_views_2020, wu_instructor_2022, wan_characterizing_2020, Stang2014Interactions}. Labs and recitations are important places for active learning to occur since they are typically smaller than the lecture courses, with more opportunities for direct interactions with the instructor and peers. These course sections are often taught by graduate teaching assistants (TAs).

Labs are often ideal places to implement active learning, because students are already expected to work in small groups on hands-on activities. Recitations are also great places for active learning because one of the major perceived barriers to implementing active learning and other research-based techniques is the perception that it interferes with the ability to cover course content. Since the supplementary sessions are in addition to lecture and not instead of it, this issue is alleviated.

Building off of work by West \textit{et al.}~\cite{West2013Variation} who investigated instructor-student interactions in a long-standing and heavily-reformed classroom context, we examine lab and recitation classroom sections that are less drastically reformed. In this study we explore what types of interactions TAs are having with students in lab and recitation environments. We compare our observations to those conducted by West \textit{et al.}~and find that while all course environments display a range of pedagogical variation, some learning environments do so more than others, and furthermore, there are certain defining features present for each environment.

\section{Research Questions}
\label{sec:guidingqs}

Our reasons for performing this study were two-fold. Practically, we were interested in collecting data that would help instructors and other academic planners at this institution make sense of the current state of affairs in the courses we observed in order to make thoughtful plans towards further reform efforts. Intellectually, we were interested in understanding the nature of instructor-student interactions in traditional labs and recitations. 

The following research questions guided our investigation:
	\begin{enumerate}
		\item What interactions are currently happening between TAs and students in traditional Labs and Recitations?
		\item How do interactions in traditional labs and recitations compare to those in more active-learning focused lab environments?
		\item Does each environment have interaction patterns that can be described as salient?
	\end{enumerate}

\section{Participants and Environment}
\label{sec:sample}

This study was conducted in an introductory physics course at a four-year, public, master's granting university. The physics department at this institution was engaged in the preliminary stages of adopting active learning pedagogies in its algebra-based introductory physics series. While the intent of the study was to examine all introductory course sessions taught by TAs, scheduling made it difficult to observe a representative sample for each course, therefore this study focuses on observations made in recitations and labs belonging to a single introductory physics course in the calculus-based introductory series which covered mechanics. The students in this course are primarily engineering and physics majors.

\subsection{Recitations}

The recitation component of the course we observed was technically optional but so strongly encouraged that the vast majority of the students in the course enrolled at the time the data was collected. Recitations were added approximately eight years prior to the data collected in this study in order to provide students an opportunity to work in small groups to solve end-of-chapter style problems. The recitation sections were discontinued in 2020 and have not since been restarted.

Recitations met once a week for one hour and fifty minutes. Students worked in small groups to complete a packet of end-of-chapter-style physics problems. The original intention was that students work together collaboratively to solve these problems. The role of the TA in recitation is that of a facilitator: to help students solve the problem sets through hints and guiding questions. The TAs are provided with solutions to the problem sets, but student work is not graded. The TA is supposed to note if individuals and groups are making an effort and are progressing towards a correct solution. Students earn a passing grade if they show up and display an effort. Recitations are limited to 24 students per section.

\subsection{Labs and Problem-Solving Sessions}

Each lab session meets once a week for two hours and fifty minutes. In lab, students work in small groups to complete an experiment and report their findings. Labs are taught by either faculty or TAs. The lab instructor is responsible for demonstrating the materials to be used as needed, troubleshooting student difficulties and finicky equipment, and grading. 

Students in lab sections follow a prescribed procedure to achieve a result and then compare results with the theoretical values. Laboratory sections are limited to 20 students per section.

As is common at other universities, students register for separate lecture and lab sections. A given lab section can therefore have students from multiple lecture sections taught by multiple lecture instructors. In the interest of consistency, every lab section of the course conducts the same experiment each week. Thus, some students entering their lab course on a given day may not have seen the material in their lecture section while other students may have. Therefore, it is the TA's responsibility to describe the requisite material for the day's activity as well as teach the students how to safely use the equipment to make adequate measurements. The TAs are also responsible for teaching sufficient data analysis methods for the students to make sense of the data they collect. This is often taught directly, in the form of a short lecture at the beginning of the lab.

In recent years,``problem solving sessions'' have been added to the lab sessions in order to improve student outcomes. During roughly one-third of the lab meetings over the course of the semester, students are given a packet of physics problems to solve instead of completing a laboratory exercise. These problem solving sessions are meant to be run in a similar manner to the Recitation sessions. Students work through a packet of end-of-chapter-style problems with their usual lab group. Typically, one or two problems in the set are turned in to be graded while the rest are offered simply as additional practice.

\subsection{Teaching Assistant Preparation}

In order to teach a Lab or Recitation section, prospective graduate student TAs must complete an application with the department. They must also be a current Masters student in the Department of Physics and Astronomy with classified standing, have an overall GPA of 3.0 or higher, and be enrolled in three units or more at SJSU. The department strives to assign all interested graduate students a teaching assignment so that graduate students may receive tuition waivers in addition to hourly wages. TAs are assigned sections to teach according to departmental need and their own course schedule, at the discretion of the department chair. Prior to the first day of classes, TAs in this study attend a three-hour training session that serves as a brief overview of physics education research findings regarding pedagogy. During the training, facilitators emphasize the importance of asking the student questions about their reasoning rather than simply providing students with solutions or merely approving correct answers. TAs engage in discussions and activities that focus on the benefits of activating prior knowledge, stating assumptions, and conceptual sense-making in general. TAs also receive an additional one-time two-hour training. This training outlines an overview of logistics of the course and TA responsibilities. For example, TAs learn where to find the information for the course, how to add students to the class, and discuss content of the activity for the first day. 
	
The TAs observed in this study were in their first or second year of graduate school. For the most part, their teaching experience was limited to tutoring and previous Lab or Recitation teaching at the same institution. The TAs observed in this study did not have weekly content prep meetings with other instructors or additional pedagogical training. However, the instructors in the CLASP study by West \textit{et al.}~\cite{West2013Variation} (who we compare our data to later in this paper) had both of these supports and a longer pre-semester training. During the semester, TAs are observed at least once by the professor in charge of a given course's Lab or Recitation sections and provided with written feedback about their teaching.

\section{Methods}		
	
\subsection{Real-time Instructor Observational Tool (RIOT)}

Observing the pedagogical moves of physics TAs and faculty instructors is hardly a new practice. Education research has yielded a number of tools for measuring the degree to which a course or instructor utilizes interactive and student-centered techniques~\cite{rtop, copus, tdop, lopus, Madsen2019}. Since one of our purposes was to compare traditional environments to the CLASP environments~\cite{Potter:Sixteen} studied by West \textit{et al.}~\cite{West2013Variation}, we chose to use the Real-time Instructor Observational Tool (RIOT). Since we were observing Labs and Recitations, we needed a tool that accounted for and categorized small-group time somehow. Furthermore, while other tools are able to indicate whether or not certain reformed techniques were being utilized, they couldn't necessarily indicate what was happening when reformed techniques were not being utilized. We chose the RIOT so that we could have a complete, continuous picture of what the TAs were doing with their time in Labs and Recitations. Using the RIOT also allows us to directly compare our results to the CLASP (Collaborative Learning through Active Sense-making in Physics) active learning curriculum observations.

Researchers at UC Davis originally developed the RIOT as a computer-based tool for monitoring and classifying instructor-student interactions~\cite{West2013Variation}. Later transformed into a free publicly-available web-based app at San Jos\'{e} State University, the RIOT quantifies and times instructor-student interactions, providing an illustrative view of what kinds of interactions actually take place in a given instructor's classroom. The RIOT enables an observer to categorize instructor actions continuously in real time. Interactions are categorized into four major groups: talking at students, shared instructor-student dialogue, observing students, and not interacting. Each of these major categories contains useful subdivisions of several descriptive categories as shown on \hyperref[tab:riot]{Table \ref*{tab:riot}}, adapted from West \textit{et al.}~\cite{West2013Variation} The instructor can perform each subcategory of interaction with an individual student, a small group of students, or with the whole class. Interaction-type, start time of the interaction, and duration of the interaction are all logged in real time. Immediately following an observation, RIOT automatically generates charts and plots that provide a complete picture of instructor-student interactions over the duration of the class period.

\newcommand{\leftcol}{0.2\textwidth-2\tabcolsep}
\newcommand{\midcol}{0.24\textwidth-2\tabcolsep}
\newcommand{\rightcol}{\textwidth-\leftcol-\midcol+2\tabcolsep}

\begin{table*}
	\centering
	\begin{tabular}{lll}
		\hline\hline
		\parbox[t]{\leftcol}{Type of \\Interaction}	&	\parbox[c][24pt][c]{\midcol}{Category of Interaction \hfill\;}	&	\parbox[c][24pt][c]{\rightcol}{Description (Instructor is\ldots) \hfill\;}	\\\hline
		\multirow{2}{*}{\parbox[t]{\leftcol}{Talking \\At Students}}	&	Clarifying Instructions	&	\parbox{\rightcol}{Clarifying the instructions, reading from the activity sheet, covering \hfill\; logistical issues, transitioning, \ldots. \hfill\;}	\\\cline{2-3}
		&	Explaining Content	&	Explaining physics concepts, answers, or processes to student(s).
			\\\hline	
			
		\multirow{4}{*}{\parbox[t]{\leftcol}{Dialoguing with Students}}	&	Listening to Question	&	Listening to a student's question.\\\cline{2-3}
			&	\parbox{\midcol}{Engaging in Closed Dialogue}	&	\parbox{\rightcol}{Asking~a~series~of~short~questions~meant~to~lead~the~student~to~a~correct \hfill\; answer. Student contribution is one to several words at a time. \hfill\;}\\\cline{2-3}
			&	Engaging in Open Dialogue	&	\parbox{\rightcol}{Students are contributing complete sentences, though not actively \hfill\; ``making sense.'' \hfill\;}\\\cline{2-3}
			&	Ideas being shared	&	\parbox{\rightcol}{Participating in student-led conversation. Student contribution is \hfill\; complete sentences with concepts being challenged and worked on. \hfill\;}\\\hline
			
		\multirow{4}{*}{\parbox[t]{\leftcol}{Observing Students}}	&	Passive Observing	&	\parbox{\rightcol}{Scanning room and assessing student progress from afar or browsing \hfill\; whiteboard work of groups for less than ten~seconds at a time. \hfill\;}\\\cline{2-3}
			&	Active Observing	&	Actively listening to small groups or individuals.\\\cline{2-3}
			&	Students Presenting	&	Listening to students presenting their work to the class.\\\cline{2-3}
			&	Students Talking Serially	&	\parbox{\rightcol}{Listening to students talking serially, asking each other questions and building on each others' ideas. \hfill\;}
			\\\hline
			
		\multirow{4}{*}{\parbox[t]{\leftcol}{Not \\Interacting}}	&	\parbox{\midcol}{Administrative and/or \hfill\; Grading \hfill\;}	&	\parbox{\rightcol}{Grading student homework, or discussing quizzes or other course \hfill\; policies. \hfill\;}\\\cline{2-3}
			&	\parbox{\midcol}{Class Preparation or \hfill\; Reading TA Notes \hfill\;}	&	Reading notes, or writing something on the board.
			\\\cline{2-3}
			&	Chatting	&	\parbox{\rightcol}{Chatting socially with students. This is not an interaction concerning physics. \hfill\;}\\\cline{2-3}
			&	\parbox{\midcol}{Working on Apparatus \hfill\; and/or Material \hfill\;}	&	\parbox{\rightcol}{Helping students with experimental apparatus or computers. Any \hfill\; possible discussion is devoid of any physics content. \hfill\;}\\\cline{2-3}
			&	Out of room	&	Left the room.\\
			\hline\hline
	\end{tabular}
	\caption[A list of Instructor-Student interactions captured by the RIOT]{A list of all possible Instructor-Student interactions captured by the RIOT, adapted from West, \emph{et al}.~\cite{West2013Variation}}
	\label{tab:riot}
\end{table*}

\subsection{Observations}

Observations took place throughout the spring semester of 2015 during the Labs and Recitations of an introductory physics course. Observations were made during the first hour of the Lab meeting or Recitation. Researchers used the RIOT as the observation protocol for all observations in this study~\cite{West2013Variation}. During the classroom observation, the observer would enter the classroom and sit in the corner with a tablet, and log the TA's interactions using the observation protocol. The TAs knew ahead of time that the observer was coming, and in some cases told the students about the observer, but sometimes not. Sometimes the students asked the observer directly why they were there, to which the observer would respond that they were observing the TA's actions. During some sessions, a quiz was administered at the beginning of lab. On those occasions, observer would wait until the end of the quiz to begin the one-hour observation. The quiz was not included in the analysis because it was not considered part of the instructional activities. 

Of the eight physics TAs employed by the department, five were observed. There were a total of sixteen laboratory or recitation sections taught by TAs of which nine were observed. Each instructor-class pair was observed at least twice. Eight sections of the lab were taught by a total of five TAs. We observed three of these TAs teaching four of these sections. Five sections of the Recitation were taught by four TAs. We observed three of these TAs teaching three of these sections. Only one TA was observed in both Lab and Recitation. A breakdown of these numbers is shown in \hyperref[tab:obs_counts]{Table~\ref*{tab:obs_counts}}.

\begin{table*}[hbt]
	\centering
	\begin{tabular}{rccccccc}
		\hline\hline
		& \multicolumn{2}{c}{Lab} & \multicolumn{2}{c}{Recitation} & \multicolumn{2}{c}{Total} \\ 
		& \parbox{2cm}{Number Observed} & \parbox{2.5cm}{Number Offered By Department} & \parbox{2cm}{Number Observed} & \parbox{2.5cm}{Number Offered By Department} & \parbox{2cm}{Number Observed} & \parbox{2.5cm}{Number Offered By Department} \\ 
		\hline
		TAs & 3 & 5 & 3 & 4 & 5 & 7 \\ 
		Sections & 4 & 8 & 3 & 5 & 5 & 13 \\ 
		\hline\hline
	\end{tabular}
	\caption{Observation counts by TA and course. To achieve a representative sample, the majority of the TAs employed by the department were observed as well as the majority of the available TA-led Lab and Recitation sections offered. One TA was observed in both Lab and Recitation.}
	\label{tab:obs_counts}
\end{table*}

\subsection{Validity of measurements}

Before beginning official observations, two researchers observed three classes as a pair. After each of the three sessions, the pair discussed the differences in observations at length resulting good agreement for our third observation. After the third session, for each second observed, our observations showed an agreement of 88.5\%. Accounting for chance agreements, we saw a Cohen's kappa value of 82.5\%~\cite{cohen_1960_coefficient}.

\subsubsection{Cohen's kappa}
	
We wrote a Python script to compare observations of a single class session collected by two different observers. For each observed interaction, the RIOT notes the start time and a duration, each with one second resolution. This is a nice feature because it means that two observations are easily synchronized by the timestamp that originates from a single clock embedded in the RIOT application. Our script compares the two observations to find temporal overlap in the two. The script reviews every second of the observations and constructs a matrix of what interaction each observer marked for every second, an example of which is shown in \hyperref[tab:kappa_matrix]{Table~\ref*{tab:kappa_matrix}}. 
\begin{table}[ht]
	\centering
	\begin{tabular}{cr|cccccc}
	\hline\hline
	& & \multicolumn{6}{c}{Observer A} \\
	& & {\rotatebox[origin=l]{90}{Clarifying}} & {\rotatebox[origin=l]{90}{Explaining}} & {\rotatebox[origin=l]{90}{Listening}} & {\rotatebox[origin=l]{90}{Closed Dialogue}} & $\cdots$ & {\rotatebox[origin=l]{90}{Not Interacting}} \\
	\hline
	\multirow{6}{*}{\rotatebox[origin=c]{90}{Observer B}}& Clarifying & 253 & 3 & 12 & 0 & $\cdots$ & 22 \\
	& Explaining & 5 & 658 & 0 & 12 & $\cdots$ & 6 \\
	& Listening & 21 & 48 & 70 & 5 & $\cdots$ & 1 \\
	& Closed Dialogue & 120 & 61 & 11 & 231 & $\cdots$ & 0 \\
	& $\vdots$ & $\vdots$ & $\vdots$ & $\vdots$ & $\vdots$ & $\ddots$ & $\vdots$ \\
	& Not Interacting & 20 & 0 & 23 & 1 & $\cdots$ & 1647 \\
	\hline\hline
	\end{tabular}
	\caption[A sample matrix used to calculate Cohen's kappa]{A sample matrix used to calculate Cohen's kappa. The ``22'' value in the top right indicates that Observer A selected \emph{Not Interacting} while Observer B had selected \emph{Clarifying Instruction} a total of twenty-two times during the observation in question.}
	\label{tab:kappa_matrix}
\end{table}
The diagonal of this matrix shows when the observations agree, with off-diagonals showing disagreement. Thus, to determine the number of agreements between two observations for the $i$th interaction type, one would simply examine the value of the $i$th diagonal.

\begin{align}
	N_{\mathrm{agreement},i} = a_{ii}
	\label{eq:agree}
\end{align}

The percent agreement is then this value divided by the sum of all the values in the matrix. Cohen's kappa goes a step further to exclude from the agreement any paired observations that may occur due to chance. To determine the number of agreements that may occur due to chance, one would compute the product of total number of times each observer marked a given category, summing over all categories. Dividing this sum by the total of all values in the matrix yields the number of chance agreements for the $i$th interaction.

\begin{align}
	N_{\mathrm{chance},i} = \frac{\sum\limits_j^n a_{ij}\sum\limits_j^n a_{ji}}{\sum\limits_j^n\sum\limits_k^n a_{jk}}
	\label{eq:chance}
\end{align}

Cohen's kappa is then determined by the difference of the numbers of agreement and chance divided by the difference of the total number of interactions and the number of chance agreements.

\begin{align}
	\kappa = \sum\limits_i^n \frac{N_{\mathrm{agreement},i} - N_{\mathrm{chance},i}}{\sum\limits_j^n\sum\limits_k^n a_{jk} - N_{\mathrm{chance},i}}
	\label{eq:kappa}
\end{align}

Values for Cohen's kappa range from $-1$ to 1, with 1 indicating perfect agreement and a value of 0 indicating chance agreement. Guidelines have been suggested to categorize $\kappa$ values from 0.61 to 0.80 as ``substantial'' and 0.81 to 1 as almost ``perfect agreement''~\cite{landisandkoch}. These guidelines have been accepted and adopted by the research community~\cite{wilcox, VonKorff, copus}. 
Another guideline characterizes $\kappa$ values over 0.75 as ``excellent'' and 0.40 to 0.75 as ``fair to good''~\cite{fleiss}. By either of these standards, the value of $\kappa = 0.825$ that we achieved is more than acceptable.

\section{Lab and Recitation Descriptive Analysis}

First we wanted to examine the types of interactions taking place in labs and recitations. \hyperref[fig:50Pies]{Figure~\ref*{fig:50Pies}} represents eight hours of observations of Labs, and \hyperref[fig:50WPies]{Figure~\ref*{fig:50WPies}} represents seven hours of observations of Recitations. We examine them both separately and then make some comparisons between the two.

\subsection{Lab}

\begin{figure*}[tbh]
	\centering
	\begin{subfigure}[m]{0.3\textwidth}
		\begin{tikzpicture}
		\tikzset{lines/.style={draw=none},}
		\pie[sum=auto, style={lines}, hide number, rotate=90, radius=1.8, color={colorNotInteracting, colorActiveObs, colorPassiveObs, colorODialogue, colorCDialogue, colorListening, colorExplaining, colorClarifying}] {60.5/60.5\%, 1.3/, 1.5/, 0.7/, 3.9/3.9\%, 4.2/4.2\%, 11.9/11.9\%, 16/16.0\%}
		\end{tikzpicture}
		\caption{Time Spent by Interaction (\unit[467.6]{min})}
		\label{fig:50TimeByInteraction}
	\end{subfigure}
	\begin{subfigure}[m]{0.3\textwidth}
		\begin{tikzpicture}
		\tikzset{lines/.style={draw=none},}
		\pie[sum=auto, style={lines}, hide number, rotate=90, radius=1.8, color={colorNotInteracting, colorSG, colorInd, colorWC}] {60.5/60.5\%, 14.7/14.7\%, 9.7/9.7\%, 15.0/15.0\%}
		\end{tikzpicture}
		\caption{{Time Spent interacting with the whole class, individuals, small groups, or no one (\unit[467.6]{min})}}
		\label{fig:50TimeByGroup}
	\end{subfigure}
	\begin{subfigure}[m]{0.3\textwidth}
		\begin{tikzpicture}[framed]
			\draw[draw=black,fill=colorNotInteracting] (0,0) rectangle (0.4,0.2) node[pos=0.5,outer sep=10pt,anchor=west] {Not Interacting};
			\draw[draw=black,fill=colorSG] (0,0.4) rectangle (0.4,0.6) node[pos=0.5,outer sep=10pt,anchor=west] {Small Groups};
			\draw[draw=black,fill=colorInd] (0,0.8) rectangle (0.4,1) node[pos=0.5,outer sep=10pt,anchor=west] {Individual};
			\draw[draw=black,fill=colorWC] (0,1.2) rectangle (0.4,1.4) node[pos=0.5,outer sep=10pt,anchor=west] {Whole Class};
		\end{tikzpicture}
	\end{subfigure}
	
	\begin{subfigure}[m]{0.3\textwidth}
		\begin{tikzpicture}
		\tikzset{lines/.style={draw=none},}
		\pie[sum=auto, style={lines}, hide number, rotate=90, radius=1.8, color={colorNotInteracting, colorActiveObs, colorPassiveObs, colorODialogue, colorCDialogue, colorListening, colorExplaining, colorClarifying}] {69.6/69.6\%, 1.5/, 1.7/, 0.8/, 4.5/4.5\%, 4.8/4.8\%, 5.3/5.3\%, 11.8/11.8\%}
		\end{tikzpicture}
		\caption{Small Group Time (\unit[406.3]{min})}
		\label{fig:50SGTime}
	\end{subfigure}
	\begin{subfigure}[m]{0.3\textwidth}
		\begin{tikzpicture}
		\tikzset{lines/.style={draw=none},}
		\pie[sum=auto, style={lines}, hide number, rotate=90, radius=1.8, color={colorCDialogue, colorListening, colorExplaining, colorClarifying}] {0.2/, 0.1/, 55.4/55.4\%, 44.3/44.3\%}
		\end{tikzpicture}
		\caption{{Whole Class Time (\unit[61.3]{min})}}
		\label{fig:50WCTime}
	\end{subfigure}
	\begin{subfigure}[m]{0.3\textwidth}
		\begin{tikzpicture}[framed]
			\draw[draw=black,fill=colorNotInteracting] (0,0.4) rectangle (0.4,0.6) node[pos=0.5,outer sep=10pt,anchor=west] {Not Interacting};
			\draw[draw=black,fill=colorPassiveObs] (0,0.8) rectangle (0.4,1) node[pos=0.5,outer sep=10pt,anchor=west] {Passive Observing};
			\draw[draw=black,fill=colorActiveObs] (0,1.2) rectangle (0.4,1.4) node[pos=0.5,outer sep=10pt,anchor=west] {Active Observing};
			\draw[draw=black,fill=colorODialogue] (0,1.6) rectangle (0.4,1.8) node[pos=0.5,outer sep=10pt,anchor=west] {Open Dialogue};
			\draw[draw=black,fill=colorCDialogue] (0,2) rectangle (0.4,2.2) node[pos=0.5,outer sep=10pt,anchor=west] {Closed Dialogue};
			\draw[draw=black,fill=colorListening] (0,2.4) rectangle (0.4,2.6) node[pos=0.5,outer sep=10pt,anchor=west] {Listening to Question};
			\draw[draw=black,fill=colorExplaining] (0,2.8) rectangle (0.4,3) node[pos=0.5,outer sep=10pt,anchor=west] {Explaining Content};
			\draw[draw=black,fill=colorClarifying] (0,3.2) rectangle (0.4,3.4) node[pos=0.5,outer sep=10pt,anchor=west] {Clarifying Instruction};
		\end{tikzpicture}
	\end{subfigure}
	\caption[TA-student interactions in the Lab classroom]{The net result of over eight hours of observations in the Lab classroom. The percentages listed are the totals of all observations rather than averages.}
	\label{fig:50Pies}
\end{figure*}
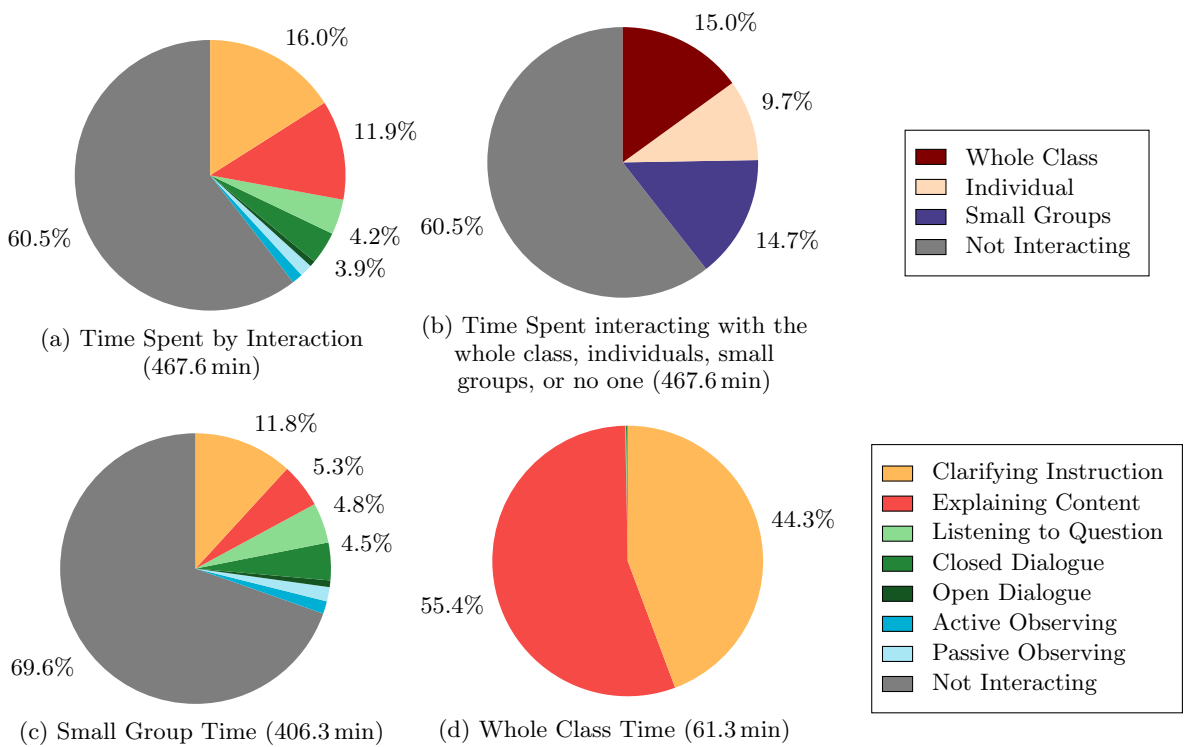

We made a total of eight observations in the Lab classes, visiting three different TAs as shown in \hyperref[tab:obs_counts]{Table~\ref*{tab:obs_counts}}. How TAs allocated their lab time is broken down by interaction in \hyperref[fig:50TimeByInteraction]{Figure~\ref*{fig:50TimeByInteraction}}. We find that TAs spent nearly 28\% of all the observations ``talking at'' students (either \emph{Clarifying Instruction} or \emph{Explaining Content}) while spending just over 8\% of the observations engaged in dialogue with students and even less time observing them. TAs spent the majority of their time being observed \emph{Not Interacting} with their students.

How TAs allocated their lab time is further broken down by group in \hyperref[fig:50TimeByGroup]{Figure~\ref*{fig:50TimeByGroup}}. We find 15\% of the time the TA was interacting with the whole class, and 9.7\% of the time was spent interacting with individual students. These findings were not surprising in that we expected TAs to spend some time addressing the class at the beginning of each lab and that it is common for students to leave their group to ask the TA a question during the lab. However, we were surprised that during the remainder of the time that students were working in small groups, so little time was spent interacting with students at all. TAs spent 60.5\% of total observation time \emph{Not Interacting} with their students. We should note here that the RIOT considers \emph{Observing} an interaction. This means that during 60.5\% of our time spent collecting data with RIOT, the attention of the TAs was somewhere other than on their students. Another way to look at this is shown in \hyperref[fig:50WCTime]{Figure~\ref*{fig:50SGTime}}. While students are working in groups, TAs are not paying attention to them nearly 70\% of that time.

All of the lab TAs addressed the whole class at the beginning of the lab. When the TA addresses the whole class during lab, this takes the form of a mini lecture that covers content related to the lab and instructions for setting up the experiment. \hyperref[fig:50WCTime]{Figure~\ref*{fig:50WCTime}} shows that there is almost no time devoted to student voices during the whole class time, as it amounts to less than one minute over all eight observations.

\subsection{Recitation}

\begin{figure*}[ht]
	\centering
	\begin{subfigure}[m]{0.3\textwidth}
		\begin{tikzpicture}
		\tikzset{lines/.style={draw=none},}
		\pie[sum=auto, style={lines}, hide number, rotate=90, radius=1.8, color={colorNotInteracting, colorIdeas, colorActiveObs, colorPassiveObs, colorODialogue, colorCDialogue, colorListening, colorExplaining, colorClarifying}] {29.76/29.8\%, 0.22/, 2.95/3.0\%/, 0.16/, 1.4/, 20.57/20.6\%, 3.27/3.3\%, 33.36/33.4\%, 8.31/8.3\%}
		\end{tikzpicture}
		\caption{Time Spent by Interaction (\unit[411.6]{min})}
		\label{fig:50WTimeByInteraction}
	\end{subfigure}
	\begin{subfigure}[m]{0.3\textwidth}
		\begin{tikzpicture}
		\tikzset{lines/.style={draw=none},}
		\pie[sum=auto, style={lines}, hide number, rotate=90, radius=1.8, color={colorNotInteracting, colorSG, colorInd, colorWC}] {29.76/29.8\%, 26.03/26.0\%, 20.87/20.9\%, 23.34/23.3\%}
		\end{tikzpicture}
		\caption{{Time Spent by Group (\unit[411.5]{min})}}
		\label{fig:50WTimeByGroup}
	\end{subfigure}
	\begin{subfigure}[m]{0.3\textwidth}
		\begin{tikzpicture}[framed]
			\draw[draw=black,fill=colorNotInteracting] (0,0) rectangle (0.4,0.2) node[pos=0.5,outer sep=10pt,anchor=west] {Not Interacting};
			\draw[draw=black,fill=colorSG] (0,0.4) rectangle (0.4,0.6) node[pos=0.5,outer sep=10pt,anchor=west] {Small Groups};
			\draw[draw=black,fill=colorInd] (0,0.8) rectangle (0.4,1) node[pos=0.5,outer sep=10pt,anchor=west] {Individual};
			\draw[draw=black,fill=colorWC] (0,1.2) rectangle (0.4,1.4) node[pos=0.5,outer sep=10pt,anchor=west] {Whole Class};
		\end{tikzpicture}
	\end{subfigure}
	
	\begin{subfigure}[m]{0.3\textwidth}
		\begin{tikzpicture}
		\tikzset{lines/.style={draw=none},}
		\pie[sum=auto, style={lines}, hide number, rotate=90, radius=1.8, color={colorNotInteracting, colorActiveObs, colorPassiveObs, colorODialogue, colorCDialogue, colorListening, colorExplaining, colorClarifying}] {40.6/40.6\%, 4/4.0\%, 0.2/, 1.9/, 20.4/20.4\%, 3.9/3.9\%, 23/23.0\%, 5.9/5.9\%}
		\end{tikzpicture}
		\caption{Small Group Time (\unit[301.5]{min})}
		\label{fig:50WSGTime}
	\end{subfigure}
	\begin{subfigure}[m]{0.3\textwidth}
		\begin{tikzpicture}
		\tikzset{lines/.style={draw=none},}
		\pie[sum=auto, style={lines}, hide number, rotate=90, radius=1.8, color={colorPresenting, colorCDialogue, colorListening, colorExplaining, colorClarifying}] {0.833/, 20.897/20.9\%, 1.637/1.6\%, 61.797/61.8\%, 14.836/14.8\%}
		\end{tikzpicture}
		\caption{{Whole Class Time (\unit[110.0]{min})}}
		\label{fig:50WWCTime}
	\end{subfigure}
	\begin{subfigure}[m]{0.3\textwidth}
		\begin{tikzpicture}[framed]
			\draw[draw=black,fill=colorNotInteracting] (0,0) rectangle (0.4,0.2) node[pos=0.5,outer sep=10pt,anchor=west] {Not Interacting};
			\draw[draw=black,fill=colorIdeas] (0,0.4) rectangle (0.4,0.6) node[pos=0.5,outer sep=10pt,anchor=west] {Students Presenting or Serial Discussion};
			\draw[draw=black,fill=colorPassiveObs] (0,0.8) rectangle (0.4,1) node[pos=0.5,outer sep=10pt,anchor=west] {Passive Obsering};
			\draw[draw=black,fill=colorActiveObs] (0,1.2) rectangle (0.4,1.4) node[pos=0.5,outer sep=10pt,anchor=west] {Active Observing};
			\draw[draw=black,fill=colorODialogue] (0,1.6) rectangle (0.4,1.8) node[pos=0.5,outer sep=10pt,anchor=west] {Open Dialogue};
			\draw[draw=black,fill=colorCDialogue] (0,2) rectangle (0.4,2.2) node[pos=0.5,outer sep=10pt,anchor=west] {Closed Dialogue};
			\draw[draw=black,fill=colorListening] (0,2.4) rectangle (0.4,2.6) node[pos=0.5,outer sep=10pt,anchor=west] {Listening to Question};
			\draw[draw=black,fill=colorExplaining] (0,2.8) rectangle (0.4,3) node[pos=0.5,outer sep=10pt,anchor=west] {Explaining Content};
			\draw[draw=black,fill=colorClarifying] (0,3.2) rectangle (0.4,3.4) node[pos=0.5,outer sep=10pt,anchor=west] {Clarifying Instruction};
		\end{tikzpicture}
	\end{subfigure}
	\caption[TA-student interactions in the Recitation classroom]{The net result of roughly seven hours of observations in Recitations. The percentages listed are the totals of all observations rather than averages.}
	\label{fig:50WPies}
\end{figure*}
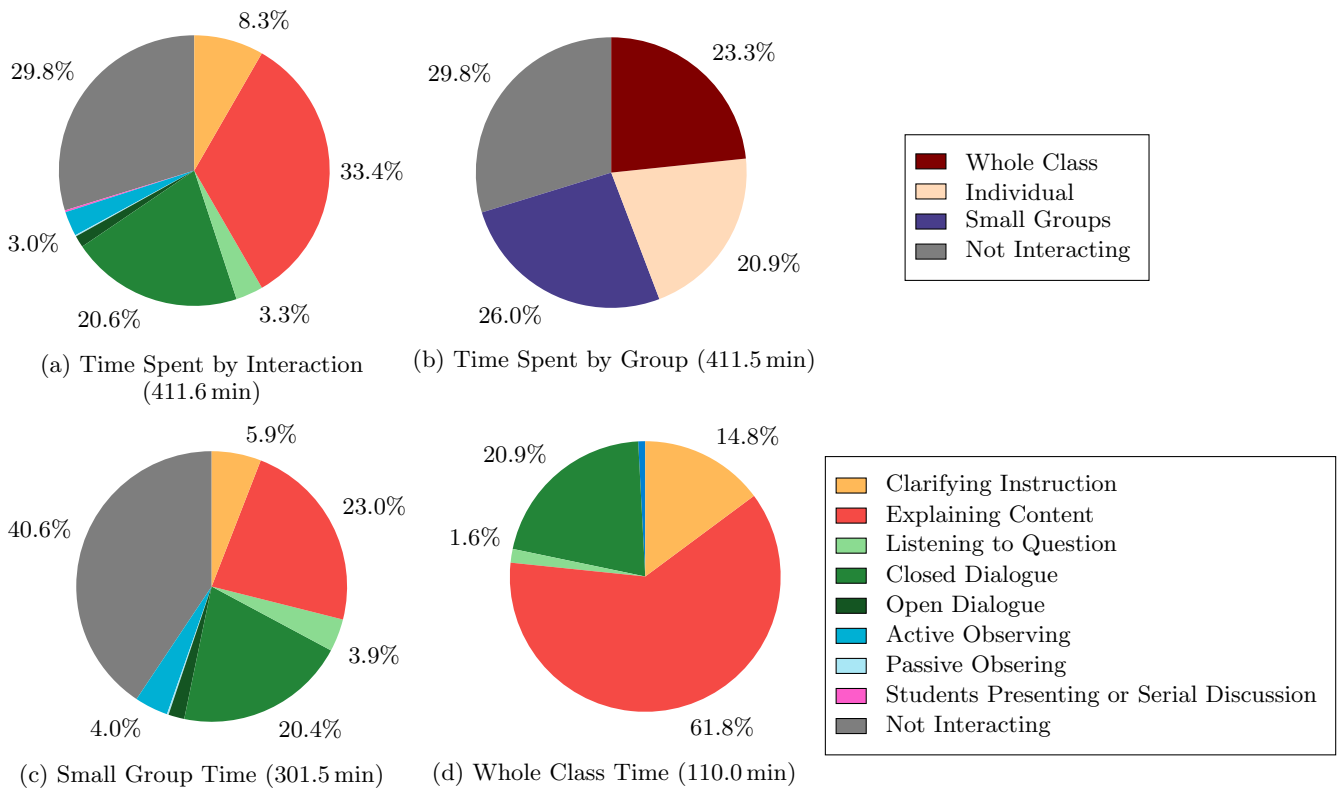

We made a total of seven observations in Recitation classes, visiting three different TAs as shown in \hyperref[tab:obs_counts]{Table~\ref*{tab:obs_counts}}. One TA was observed in both Lab and Recitation. Students in Recitations saw a different landscape of interactions than those in the Lab class. \hyperref[fig:50WTimeByGroup]{Figure~\ref*{fig:50WTimeByGroup}} shows that TAs split their time nearly evenly between working with the whole class, individual students, small groups, and not interacting. The \emph{Not Interacting} portion is 29.8\% of the total time, followed by \emph{Small Groups}, \emph{Whole Class}, and \emph{Individual} at 26.0\%, 23.3\%, and 20.9\%, respectively. The data summarized by the charts in \hyperref[fig:50WPies]{Figure~\ref*{fig:50WPies}}, show that Recitation TAs spent more time talking with their students in the form of \emph{Open} or \emph{Closed Dialogue} than the lab TAs did.

Recitations are more interactive than labs in that only 29.8\% of total time is spent \emph{Not Interacting}. This is not entirely surprising given that the nature of lab often requires students to set up experiments, collect data, and make sense of these data alone or in pairs, whereas the recitations are intended to be an opportunity for students to practice their problem solving and get feedback from instructors. However, when the students are working in small groups during recitation, 40.6\% of the time instructors are still not paying attention to them.

During the Recitations, a large percentage of total time (33.4\%) is spent \emph{Explaining Content}. Since 29.8\% of total time is spent \emph{Not Interacting} with students at all, approximately half of all time TAs spend interacting with students is spent \emph{Explaining Content} to them.

Interestingly, we also find that the interaction profile of our Recitations is very similar to the average interaction profile of the mini studios examined in by Wilcox \textit{et al.}~\cite{wilcox_quicker_2016} The sections they examined had active learning elements, however, the authors indicated that the curriculum was not implemented with fidelity, citing that the instructors explained and clarified more than the curriculum intended, and that the instructors spent more time not interacting than ideal. 

\subsection{Comparison between Lab and Recitation classrooms}

\begin{figure*}[htb]
	\centering
	\begin{subfigure}[m]{0.32\linewidth}
	\begin{tikzpicture}
		\pgfplotstableread[col sep=comma]{Data/SG+WCData-50-transposed.dat}{\csvdata}
		\begin{axis}[
			width=1.1*\textwidth, 
			font=\tiny,
			boxplot/draw direction=y,
			ymin=0,ymax=100,ytick={0,20,...,100},
			ymajorgrids=true,
			minor y tick num=1,
			yminorgrids=true,
			ylabel={Percent of Observation},
			ylabel style={overlay=true, at={(0.125,0.5)}},
			yticklabel style={overlay=true},
			major grid style={dashed},
			minor grid style={dotted},
			xtick={1,...,8},
			xtick style={draw=none},
			xticklabel style={align=right,rotate=90},
			xticklabels={Clarifying, Explaining, Listening, Closed\\Dialogue, Open\\Dialogue, Observing, Students\\Presenting, Not\\Interacting},
			cycle list name=SgWcColorList,
		]
		\foreach \n in {0,...,7} {
			\addplot+ [boxplot,draw=black,fill] table[y index=\n] {\csvdata};
		}
		\end{axis}
		\end{tikzpicture}
		\caption{Lab}
		\label{fig:SJSUlab}
	\end{subfigure}
	\begin{subfigure}[m]{0.32\linewidth}
		\begin{tikzpicture}
		\pgfplotstableread[col sep=comma]{Data/SG+WCData-50W-transposed.dat}{\csvdata}
		\begin{axis}[
			width=1.1*\textwidth,
			font=\tiny,
			boxplot/draw direction=y,
			ymin=0,ymax=100,ytick={0,20,...,100},
			ymajorgrids=true,
			minor y tick num=1,
			yminorgrids=true,
			ylabel={Percent of Observation},
			ylabel style={overlay=true, at={(0.125,0.5)}},
			yticklabel style={overlay=true},
			major grid style={dashed},
			minor grid style={dotted},
			xtick={1,...,8},
			xtick style = {draw=none},
			xticklabel style={align=right,rotate=90},
			xticklabels={Clarifying, Explaining, Listening, Closed\\Dialogue, Open\\Dialogue, Observing, Students\\Presenting, Not\\Interacting},
			cycle list name=SgWcColorList,
		]
		\foreach \n in {0,...,7} {
			\addplot+ [boxplot,fill,draw=black] table[y index=\n] {\csvdata};
			}
		\end{axis}
		\end{tikzpicture}
		\caption{Recitation}
		\label{fig:SJSUrecitation}
	\end{subfigure}
	\begin{subfigure}[m]{0.32\linewidth}
		\begin{tikzpicture}
		\pgfplotstableread[col sep=comma]{Data/SG+WCData-CLASP-transposed.dat}{\csvdata}
		\begin{axis}[
			width=1.1*\textwidth,
			font=\tiny,
			boxplot/draw direction=y,
			ymin=0,ymax=100,ytick={0,20,...,100},
			ymajorgrids=true,
			minor y tick num=1,
			yminorgrids=true,
			ylabel={Percent of Observation},
			ylabel style={overlay=true, at={(0.125,0.5)}},
			yticklabel style={overlay=true},
			major grid style={dashed},
			minor grid style={dotted},
			xtick={1,...,8},
			xtick style = {draw=none},
			xticklabel style={align=right,rotate=90},
			xticklabels={Clarifying, Explaining, Listening, Closed\\Dialogue, Open\\Dialogue, Observing, Students\\Presenting, Not\\Interacting},
			cycle list name=SgWcColorList,
		]
		\foreach \n in {0,...,7} {
			\addplot+ [boxplot,fill,draw=black] table[y index=\n] {\csvdata};
			}
		\end{axis}
		\end{tikzpicture}
		\caption{CLASP}
		\label{fig:Davisclasp}
	\end{subfigure}
	\caption[Box-and-whisker plot comparing ``Whole Class'' time in SJSU and UC~Davis courses]{Range of interactions by fraction of total class time in Labs, Recitations, and CLASP~\cite{West2013Variation}}
	\label{fig:SJSUvsDavis}
\end{figure*}
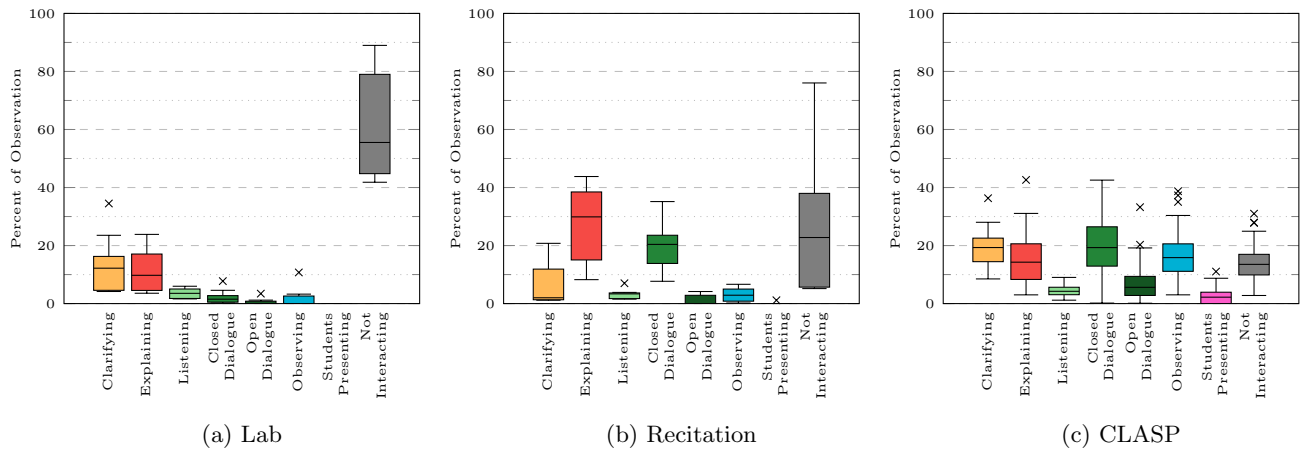

When comparing the interactions in the Labs and Recitations, we found a wider range of types of interactions for the Recitations, and also found that the ways in which TAs divided up their time in Recitation was also more widely varied than in lab, as illustrated by the larger variation of the amount of time spent engaging in each type of interaction in \hyperref[fig:SJSUrecitation]{Figure~\ref*{fig:SJSUrecitation}} compare to \hyperref[fig:SJSUlab]{Figure~\ref*{fig:SJSUlab}}. For example, while one TA spent the majority of Recitation time interacting with the ``Whole Class,'' another TA spent none, not even in the form of a greeting at the beginning of the session. Of the two TAs who did interact with the ``Whole Class,'' each time they were observed, they both started each day with a short talk discussing the topic of the day. The format of the talk varied by day and by TA (\emph{Clarify}, \emph{Explain}, and \emph{Dialogue} were all used) and ranged from two to eight minutes.

In the Recitation setting, TAs tended to spend a lot of time \emph{Not Interacting} with their students, averaging between a quarter and a third of the session. They spend even more time \emph{Not Interacting} in the lab sections, spending just under 60\% of the time \emph{Not Interacting}.

\subsection{Interpretation of Lab and Recitation Results}
\label{sec:InterpretationOfResults-LearningGoals}

One of the more informative findings we took from this analysis was the general lack of interactions between TAs and students. Taking into account the separate goals of lab and recitation, this finding has a different meaning depending on which of the learning goals are emphasized.

In the lab setting alone, this finding has two very different interpretations. 
\begin{enumerate}
 \item The nature of the lab is such that interactions with the instructor are not needed or intended to occur for the majority of the time, or
 \item The instructor could and should be interacting productively with the students to provide better support, but for any number of reasons is not. 
\end{enumerate}
Both interpretations are dependent on the learning goals of the lab and have implications for how departments allocate resources. If the main purpose of lab time is for students to get practice setting up experiments, using equipment, and collecting data by following written instructions on how to do so, those in charge of the course might decide that the level of interactions with the TA are appropriate and they then might decide to increase the enrollment size of labs, or use these data as evidence that a larger lab space could be built and facilitated by a single instructor. But if the main purpose of lab time is for students to figure how to design experiments to address a question, and/or to interpret results as consistent or inconsistent with theory, then those in charge of the class would probably agree that the students are in need of more feedback from the TA than they are currently getting. They may choose to design different curricular materials to make sure students are doing these things, and develop TA guides or professional development opportunities that help the TAs understand how they can be more supportive of these activities.

For the recitation, the implication of the large amount of non-interaction time is similarly dependant on the purpose of session. Possible objectives for the recitation could be for students
\begin{enumerate}
 \item to develop physics problem solving skills, 
 \item to do physics sense-making in an environment that supports active learning, and/or
 \item to build community around doing physics.
\end{enumerate}
Arguably, it is possible to develop physics problem solving skills without actually engaging in sense making~\cite{Kim2002Students}. In this case, the students could be expected to perform these same activities on their own time without the cost of a facilitator, or the section again could be enlarged to allow one facilitator to manage more students. However, if one of the course goals is to increase physics identity and community, it might make sense to increase the involvement from the facilitator. Therefore, given the amount of non-interaction we see in the Recitation, we expect that those in charge of the course would make adjustments to the materials, environment, and TA training and support related to this course.

\section{Comparison to CLASP data}
\label{sec:CompareToCLASP}

While some amount of non-interaction is expected, the results from the Labs and Recitations are very different from the results from West's study of the CLASP curriculum. There it is reported that TAs spend an average of 20\% of Small Group time and therefore an even smaller percentage of total time \emph{Not Interacting} with students~\cite{West2013Variation}.

There exist a few studies examining instructors interacting with their students using RIOT~\cite{West2013Variation, wilcox}. Therefore, we have an opportunity to compare our data with that from other studies.

A study at UC~Davis examined their model-based CLASP course~\cite{West2013Variation, Potter:Sixteen}. The course is their introductory physics course for non-majors. The interactive portion of the course, called ``Discussion/Lab'' (DL), has students work together in small groups to answer a series of activity prompts followed by a whole class discussion of the major points of the activity and explore general implications of the phenomena. Instructors, largely graduate students and a few faculty, are coached to guide students toward correct usage of the models rather than provide immediate answers. Effectively, the CLASP course is designed to be a more interactive hybrid of both the lab and recitation settings studied in this paper. As such, we might expect some similarities in the spreads of interaction types.

Looking at \hyperref[fig:SJSUvsDavis]{Figure~\ref*{fig:SJSUvsDavis}}, we again see only slight similarities between the Lab and Recitation and CLASP. \emph{Clarifying Instruction} shows similar ranges across the three contexts while \emph{Explaining Content} show similar ranges in CLASP and Lab, with Recitation featuring more explanation time. \emph{Closed Dialogue} in Recitation is similar to CLASP, while in Lab \emph{Closed Dialogue} is, at best, half that of CLASP while \emph{Listening to Questions} is comparable across the three contexts.

The similarities of either Lab or Recitation compared to CLASP end there. Students in both Lab and Recitation courses were never provided the opportunity to \emph{Present} their lab findings and rarely \emph{Presented} to each other or \emph{Talked Serially} to build each others' ideas. These interactions show up in the CLASP observations, as \emph{Students Presenting}. The opportunity for \emph{Student Presentations} plays a fairly important role in the CLASP material and student sense making in general. Furthermore there is almost no \emph{Observing} of students in Lab and Recitation, while in CLASP, this interaction is much more common. The largest difference between CLASP and the two other contexts is the amount of time that the instructors are \emph{Not Interacting} in Lab and Recitation. This finding is consistent with Wu \textit{et al.}~\cite{wu_instructor_2022} who also found that traditional labs did not support interactions between instructors and students.

When the TAs in the labs and recitations are interacting with their students, they tend to spend more time ``talking at'' their students than ``talking with'' them. While the recitation has a slightly higher median \emph{Closed Dialogue} percentage, \emph{Listening to Questions} is comparable across all courses. In both the lab and recitation courses, the percentage of time spent \emph{Observing} is far less than in the CLASP course. Interestingly, Recitation TAs tended to spend a higher percentage of their ``Small Group'' time \emph{Explaining Content} while the Lab TAs spent a lower percentage of their ``Small Group'' time \emph{Explaining}. CLASP TAs fall neatly in between these two groups. They did tend to spend a larger fraction of this time \emph{Clarifying Instruction} than TAs in both the lab and the recitation. This is likely due to the fact that the materials provided to the lab and recitation students are far more explicit in their instructions for students compared to the DL materials provided to the CLASP students.

Based on our observations, the Recitation course more closely aligns with the CLASP course than the Lab. However, the data show that the range of interactions in the Recitation sections was larger than that of the CLASP course, in particular considering \emph{Explaining} and \emph{Not Interacting}. This suggests that the Recitation section has the potential to be an interactive course, but might not always be so depending on how the instructor chooses to interact with the students during Recitation. One additional notable difference is that there were no \emph{Student Presentations} in the Recitation sections.

If the goal is to have a consistently actively engaged facilitator in Lab and Recitation, TAs will require significantly more training and departmental support in the form of professional development and/or curricular materials.

\subsection{Interactivity of Labs, Recitations, and CLASP}
\label{sec:Interactivity}

The RIOT provides an illustrative overview of classroom interactions, but it does not attempt to describe the content or quality of these interactions. Thus, making inferences regarding the extent to which a particular observation exemplifies best-practices can prove difficult. However, particular features of the RIOT can serve as proxies for best-practices, and/or indicate the absence of others. For example, \emph{Dialogue} between instructors and students indicates the potential for students to obtain feedback from their instructor about their thinking process. Similarly, long periods of TA \emph{Explaining} without presence of student voice indicates a lack of student opportunity to participate. It is here that we make a distinction between interactivity and active learning. Active learning encompasses a wide range of teaching practices that doesn't lend itself to a singular definition~\cite{meltzer_resource_2012}. However, in the interest of making a point about the way in which RIOT can be used to to collect evidence for active learning, we will adopt the following definition. Active learning (also known as interactive engagement) has been described as methods meeting the three criteria~\cite{hake}.
 \begin{enumerate}
    \item{Students perform activities that promote of conceptual understanding}
    \item{Students engage in active thinking (and potentially engage in a hands-on way)} 
    \item{Students are given real-time feedback by peers and/or instructors}
 \end{enumerate}
The RIOT does not tell us if the activities the students are engaging in promote conceptual understanding (the RTOP~\cite{rtop} would be a good measure of this), nor does it tell us about the nature of student-student interactions when the instructor is not interacting with them (the TDOP~\cite{tdop} and COPUS~\cite{copus} can), but the RIOT can tell us if the instructor is promoting a dialogic atmosphere among students, which allows for feedback between instructor, students, and peers. Furthermore, it gives us a breakdown of the amount of time spent ``talking at'' and ``talking with'' students. The RIOT really tells us about the nature of the facilitation (or lack there of) that is taking place in the classroom. 

If we use the RIOT to observe a class which has lots of shared student-instructor dialogue and few periods of extended explaining of content, the class has the potential to contain active learning elements, but we can only make claims about the extent to which it is interactive. This is an important distinction. 

We don't make claims that the RIOT measures active learning directly, but instead that it gives us an illustrative view of the nature of the instructor-student interactions that are taking place in the classroom which can provide evidence for or against interactivity, which allows us to make some inferences about active learning (which can also potentially be bolstered by other kinds of data collection).

Using this reasoning, we recognize that the RIOT categorizes students working in small groups without the TA as \emph{Not Interacting} for the TA even though the students may be interacting with each other and may even be engaging with each other in a way that is consistent with the definition of an active learning course. That said, if the TA is \emph{Not Interacting} with or even \emph{Observing} any students for long periods of time, the TA is not providing feedback to the students and is not modeling a dialogic environment where students' ideas are valued. For example, Wan \textit{et al.}~\cite{wan_characterizing_2020} find that when instructors use more interactive techniques, the students are more engaged in the lab environment. This provides the framework for our next analysis.

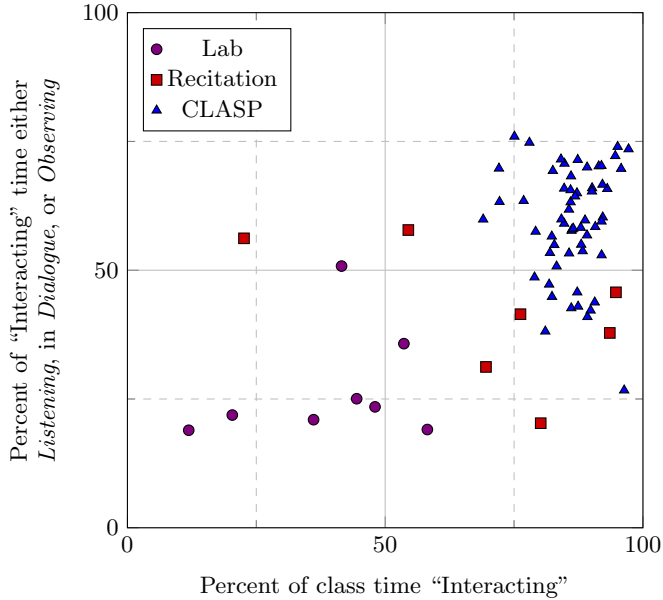
\begin{figure}[htb]
\centering
\begin{tikzpicture}
\begin{axis}[
 height=\axisdefaultwidth,
 width=\axisdefaultwidth,
 ylabel style={text width=\axisdefaultwidth,align=center},
	ylabel={Percent of ``Interacting'' time either \\\emph{Listening}, in \emph{Dialogue}, or \emph{Observing}},
	ymin=0,ymax=100,
	ytick={0,50,100},
	xlabel={Percent of class time ``Interacting''},
	xmin=0,xmax=100,
	xtick={0,50,100},
	minor tick num=1,
	grid=both,
	minor grid style={dashed},
	legend entries={Lab,Recitation,CLASP},
	legend pos=north west,
	]
	\addplot+ [
		only marks,
		point meta=explicit symbolic,
		violet,
		mark=*,
		draw=black,
		mark options={fill=violet},
	] table [meta=label] {Data/TalkingWithVSInteracting-50.dat};
	\addplot+ [
		only marks,
		point meta=explicit symbolic,
		mark=square*,
		draw=black,fill=red,
	] table [meta=label] {Data/TalkingWithVSInteracting-50W.dat};
	\addplot [
		scatter,
		only marks,
		point meta=explicit symbolic,
		scatter/classes={
			CLASP={mark=triangle*},
			Ideas={mark=triangle*}
		},
		draw=black,
		fill=blue,
	] table [meta=label] {Data/TalkingWithVSInteracting-CLASP2.dat};
\end{axis}
\end{tikzpicture}
	\caption{Percent of ``Interacting'' time spent ``Talking \emph{with}'' students as a function of percent of class time ``Interacting'' with students}
	\label{fig:TalkingWithVsInteracting}
\end{figure}

In \hyperref[fig:TalkingWithVsInteracting]{Figure~\ref*{fig:TalkingWithVsInteracting}} we plot the Recitations and Labs we observed against the CLASP sections observed by West \textit{et al.} We wanted to separate lecture and explaining physics from some of the other types of interactions that support the ellicitation of student ideas. Therefore, we plotted interaction time along the horizontal axis and the percent of that interaction time spent in \emph{Dialogue With}, \emph{Listening To}, and \emph{Observing} students vertically. We notice that all the CLASP TAs interact with their students more than 70\% of class time, while the TAs teaching the labs and recitations interact with their students anywhere from 10\% to 95\% of class time. We also notice that Lab and Recitation TAs spend on average much more of their interaction time \emph{Explaining Content} (``talking at'') to students rather than ``talking with'' them or \emph{Observing} them as they work in groups.

We imagine that traditional lectures would fall in the lower-left quadrant of this plot and that active learning courses would generally fall in the upper-right quadrant of the plot. The fact that we see the labs and recitations all over the plot, indicates to us that we need more support in the form of curricular prompts and professional development for TAs so that they can be more comfortable being the types of facilitators that the learning goals demand. The fact that TAs in Labs and Recitations are more likely to fall toward the bottom of the plot (``lectures more than listens'') indicate that the curriculum and supporting TA materials themselves may need more instruction on how to include and value student voices during investigation in order to foster a ``heads on'' experience if the goal is to include more active learning practices in those courses.

\begin{figure}[htb]
\centering
\begin{tikzpicture}
\begin{axis}[
 height=\axisdefaultwidth,
 width=\axisdefaultwidth,
 ylabel style={text width=\axisdefaultwidth,align=center},
	ylabel={Percent of ``Interacting'' time \\ either \emph{Listening} or in \emph{Dialogue}},
	ymin=0,ymax=100,
	ytick={0,50,100},
 xlabel style={text width=\axisdefaultwidth,align=center},
	xlabel={Percent of class time ``Interacting''\\(not counting \emph{Observing})},
	xmin=0,xmax=100,
	xtick={0,50,100},
	minor tick num=1,
	grid=both,
	minor grid style={dashed},
	legend entries={Lab,Recitation,CLASP},
	legend pos=north west,
	]
	\addplot+ [
		only marks,
		point meta=explicit symbolic,
		violet,
		mark=*,
		draw=black,
		mark options={fill=violet},
	] table [meta=label] {Data/TalkingWithVSInteracting-50-noObs.dat};
	\addplot+ [
		only marks,
		point meta=explicit symbolic,
		mark=square*,
		draw=black,fill=red,
	] table [meta=label] {Data/TalkingWithVSInteracting-50W-noObs.dat};
	\addplot [
		scatter,
		only marks,
		point meta=explicit symbolic,
		scatter/classes={
			CLASP={mark=triangle*},
			Ideas={mark=triangle*}
		},
		draw=black,
		fill=blue,
	] table [meta=label] {Data/TalkingWithVSInteracting-CLASP2-noObs.dat};
\end{axis}
\end{tikzpicture}
	\caption{Percent of ``Interacting'' time spent \emph{not} ``Talking at'' students as a function of percent of class time ``Interacting'' with students. Note, for this figure, we do \emph{not} include RIOT's \emph{Observing} codes as as time spent interacting with students as we did in \hyperref[fig:TalkingWithVsInteracting]{Figure~\ref*{fig:TalkingWithVsInteracting}}.}
	\label{fig:TalkingWithVsInteracting-Obs=NI}
\end{figure}
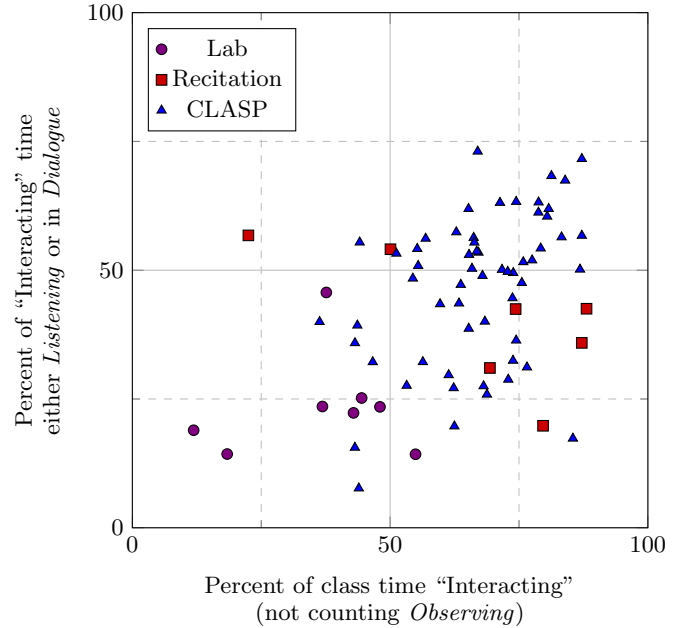

When defining interactivity, we wondered how to categorize the action of \emph{Observing}. If an instructor is \emph{Observing} a student, the student isn't necessarily participating in the interaction. Wu \textit{et al.} defined interactions as speech to or from the instructor in their 2022 study of interactions between instructors and students in the classroom~\cite{wu_instructor_2022}, which does not include \emph{Observing}. We were interested in seeing if removing \emph{Observing} from the definition of interactivity impacted our results. If we explicitly do not consider a TA \emph{Observing} their students as a method of ``Interacting,'' we arrive at \hyperref[fig:TalkingWithVsInteracting-Obs=NI]{Figure~\ref*{fig:TalkingWithVsInteracting-Obs=NI}}. Here, we see a greater mingling of the three, otherwise very different courses. There is much more overlap between the CLASP curriculum and general Recitations. Indeed, it becomes much more difficult to differentiate CLASP from the majority of our recitation observations. We conclude then that \emph{Observation} remains an important interaction to monitor between instructors and students, as it seems to be a major distinguishing factor between Recitation sections and CLASP sections.

\section{Discussion}

It is important to note what the overall results here do and don't tell us. The RIOT shows us to what extent an instructor interacts with the students in her/his classroom. However, it does not tell us to what extent this interactivity constitutes ``active learning.''

The extent to which these implementation differences have an effect on student learning outcomes is an important open area of research cited by Docktor and Mestre~\cite{Docktor2014}. 

Therefore, active learning is supported by instructors who act as facilitators of learning rather than sources of knowledge. Pedagogies associated with being a facilitator are arguably different from those of an instructor who acts as a source of knowledge. For example, an instructor acting as a facilitator of learning might spend less time explaining content than one who is acting as a source of knowledge. This is not to say that \emph{Explaining Content} is bad, or that a facilitator doesn't do any \emph{Explaining}, but rather that amount of class times devoted to \emph{Explaining Content} is likely to be different between those who consider themselves facilitators of active learning and those who see themselves as sources of knowledge.

The data presented here illuminate many areas for potential growth in labs and recitations. Future interventions and subsequent observations could easily be designed to target any specific objectives.

One major takeaway from this project is that we now have an understanding what interactions are happening in each context, and we can now think more carefully about which kinds of interactions will support the learning goals for each kind of learning environment. 

\section{Conclusions}

\subsection{A Summary of Results}

What follows is a short summary of our findings, organized by the appropriate research question. References to the original figures are included as well as a brief discussion of their implications and important points for future TA professional development.

\subsubsection{What TA-student interactions do our TAs employ in their teaching assignments?}

Combining the total times from \hyperref[fig:50Pies]{Figures~\ref*{fig:50Pies}} and \hyperref[fig:50WPies]{\ref*{fig:50WPies}} and working with the percentages, we find that the TAs spent just over 34\% of the time ``talking at'' their students. About 17\% of the total time was spent ``talking with'' or observing their students. TAs spent 45\% of the total time \emph{Not Interacting} with their students. They spent comparable amounts of time interacting with the Whole Class as they did with Small Groups and only slightly less time interacting with students Individually.

\subsubsection{How do these interactions differ in Lab and Recitation?}

\hyperref[fig:SJSUlab]{Figure~\ref*{fig:SJSUlab}} shows relatively small spreads in the interactions used by TAs teaching lab. Of the interactions, both \emph{Clarifying Instruction} and \emph{Explaining Content} were implemented the most. Both interactions also showed the largest spread in amount of time used. Except for a single outlier, \emph{Listening to Question}, \emph{Open/Closed Diagloue}, and \emph{Observing} all fall entirely in the single-digit percentages of total observation time. Even with the largest percentage spread, TAs spent the majority of their lab time \emph{Not Interacting} with their students.

Overall, variations in interactions are more likely due to variation by TA than any other factor, possibly due to similar thoughts and beliefs about how a lab course should be run.

\hyperref[fig:SJSUrecitation]{Figure~\ref*{fig:SJSUrecitation}} shows much greater variation by interaction type in Recitation compared to the Lab data in \hyperref[fig:SJSUlab]{Figure~\ref*{fig:SJSUlab}}. There appears to be less of a consensus about how to run a recitation course. This leads to varying student experiences in recitation that depend on which TA leads their recitation. In the interest of a homogenized recitation experience, more training and departmental support is needed. Here, TAs could become more interactive with useful, clear-cut examples and guidance from more experienced instructors. Observations of others teaching the same course would be beneficial to their professional development, as would time to reflect on their own teaching practices.

\subsubsection{How do our TAs' implementations of Lab and Recitations compare with available RIOT data from CLASP}

Comparing the variability of interactions across the three class settings shown in \hyperref[fig:SJSUvsDavis]{Figure~\ref*{fig:SJSUvsDavis}}, TAs in the SJSU Labs and Recitations have a stronger tendency to ``talk at'' their students than their CLASP counterparts. They also spend considerably more time \emph{Not Interacting}.

CLASP instructors also spend much more time observing than instructors from our Labs and Recitations. Because CLASP represents a carefully reformed course designed to elicit student ideas, we conclude that \emph{Observing} could be an unexpected characteristic of reformed classrooms that should be investigated for pedagogical importance.

\section{Acknowledgments}
We would like to thank Katrina Roseler, Annie Chase, Zairac Smith, and the rest of the San Jos\'e State University PER group for their support and guidance during the data collection stage of this project. The authors would also like to thank Emily West who collected half the CLASP data from the original 2013 study, and David Webb who reviewed an earlier copy of this manuscript. This work would not have been possible without the support of San Jos\'e State University’s Research, Scholarship, and Creative Activity Assigned Time Program. This work is supported in part by the now terminated National Science Foundation grant No.\ 1953760. 

\bibliography{Bibliography}

\begin{thebibliography}{34}%
\makeatletter
\providecommand \@ifxundefined [1]{%
 \@ifx{#1\undefined}
}%
\providecommand \@ifnum [1]{%
 \ifnum #1\expandafter \@firstoftwo
 \else \expandafter \@secondoftwo
 \fi
}%
\providecommand \@ifx [1]{%
 \ifx #1\expandafter \@firstoftwo
 \else \expandafter \@secondoftwo
 \fi
}%
\providecommand \natexlab [1]{#1}%
\providecommand \enquote  [1]{``#1''}%
\providecommand \bibnamefont  [1]{#1}%
\providecommand \bibfnamefont [1]{#1}%
\providecommand \citenamefont [1]{#1}%
\providecommand \href@noop [0]{\@secondoftwo}%
\providecommand \href [0]{\begingroup \@sanitize@url \@href}%
\providecommand \@href[1]{\@@startlink{#1}\@@href}%
\providecommand \@@href[1]{\endgroup#1\@@endlink}%
\providecommand \@sanitize@url [0]{\catcode `\\12\catcode `\$12\catcode
  `\&12\catcode `\#12\catcode `\^12\catcode `\_12\catcode `\%12\relax}%
\providecommand \@@startlink[1]{}%
\providecommand \@@endlink[0]{}%
\providecommand \url  [0]{\begingroup\@sanitize@url \@url }%
\providecommand \@url [1]{\endgroup\@href {#1}{\urlprefix }}%
\providecommand \urlprefix  [0]{URL }%
\providecommand \Eprint [0]{\href }%
\providecommand \doibase [0]{https://doi.org/}%
\providecommand \selectlanguage [0]{\@gobble}%
\providecommand \bibinfo  [0]{\@secondoftwo}%
\providecommand \bibfield  [0]{\@secondoftwo}%
\providecommand \translation [1]{[#1]}%
\providecommand \BibitemOpen [0]{}%
\providecommand \bibitemStop [0]{}%
\providecommand \bibitemNoStop [0]{.\EOS\space}%
\providecommand \EOS [0]{\spacefactor3000\relax}%
\providecommand \BibitemShut  [1]{\csname bibitem#1\endcsname}%
\let\auto@bib@innerbib\@empty
\bibitem [{\citenamefont {West}\ \emph {et~al.}(2013)\citenamefont {West},
  \citenamefont {Paul}, \citenamefont {Webb},\ and\ \citenamefont
  {Potter}}]{West2013Variation}%
  \BibitemOpen
  \bibfield  {author} {\bibinfo {author} {\bibfnamefont {E.~A.}\ \bibnamefont
  {West}}, \bibinfo {author} {\bibfnamefont {C.~A.}\ \bibnamefont {Paul}},
  \bibinfo {author} {\bibfnamefont {D.}~\bibnamefont {Webb}},\ and\ \bibinfo
  {author} {\bibfnamefont {W.~H.}\ \bibnamefont {Potter}},\ }\bibfield  {title}
  {\bibinfo {title} {{Variation of instructor-student interactions in an
  introductory interactive physics course}},\ }\href
  {https://doi.org/10.1103/physrevstper.9.010109} {\bibfield  {journal}
  {\bibinfo  {journal} {Phys. Rev. ST Phys. Educ. Res.}\ }\textbf {\bibinfo
  {volume} {9}},\ \bibinfo {pages} {010109+} (\bibinfo {year}
  {2013})}\BibitemShut {NoStop}%
\bibitem [{\citenamefont {Holmes}\ and\ \citenamefont
  {Wieman}(2018)}]{holmes_introductory_2018}%
  \BibitemOpen
  \bibfield  {author} {\bibinfo {author} {\bibfnamefont {N.~G.}\ \bibnamefont
  {Holmes}}\ and\ \bibinfo {author} {\bibfnamefont {C.~E.}\ \bibnamefont
  {Wieman}},\ }\bibfield  {title} {\bibinfo {title} {Introductory physics labs:
  We can do better},\ }\href {https://doi.org/10.1063/PT.3.3816} {\bibfield
  {journal} {\bibinfo  {journal} {Physics Today}\ }\textbf {\bibinfo {volume}
  {71}},\ \bibinfo {pages} {38} (\bibinfo {year} {2018})}\BibitemShut {NoStop}%
\bibitem [{\citenamefont {Hake}(1998)}]{hake}%
  \BibitemOpen
  \bibfield  {author} {\bibinfo {author} {\bibfnamefont {R.~R.}\ \bibnamefont
  {Hake}},\ }\bibfield  {title} {\bibinfo {title} {{Interactive-engagement
  versus traditional methods: A six-thousand-student survey of mechanics test
  data for introductory physics courses}},\ }\href
  {https://doi.org/10.1119/1.18809} {\bibfield  {journal} {\bibinfo  {journal}
  {American Journal of Physics}\ }\textbf {\bibinfo {volume} {66}},\ \bibinfo
  {pages} {64} (\bibinfo {year} {1998})}\BibitemShut {NoStop}%
\bibitem [{\citenamefont {Crouch}\ and\ \citenamefont {Mazur}(2001)}]{mazur}%
  \BibitemOpen
  \bibfield  {author} {\bibinfo {author} {\bibfnamefont {C.~H.}\ \bibnamefont
  {Crouch}}\ and\ \bibinfo {author} {\bibfnamefont {E.}~\bibnamefont {Mazur}},\
  }\bibfield  {title} {\bibinfo {title} {{Peer instruction: Ten years of
  experience and results}},\ }in\ \href
  {http://citeseerx.ist.psu.edu/viewdoc/summary?doi=10.1.1.113.6060} {\emph
  {\bibinfo {booktitle} {American Journal of Physics}}}\ (\bibinfo {year}
  {2001})\ pp.\ \bibinfo {pages} {970--977}\BibitemShut {NoStop}%
\bibitem [{\citenamefont {Von~Korff}\ \emph {et~al.}(2016)\citenamefont
  {Von~Korff}, \citenamefont {Archibeque}, \citenamefont {Gomez}, \citenamefont
  {Heckendorf}, \citenamefont {McKagan}, \citenamefont {Sayre}, \citenamefont
  {Schenk}, \citenamefont {Shepherd},\ and\ \citenamefont
  {Sorell}}]{VonKorff2016}%
  \BibitemOpen
  \bibfield  {author} {\bibinfo {author} {\bibfnamefont {J.}~\bibnamefont
  {Von~Korff}}, \bibinfo {author} {\bibfnamefont {B.}~\bibnamefont
  {Archibeque}}, \bibinfo {author} {\bibfnamefont {K.~A.}\ \bibnamefont
  {Gomez}}, \bibinfo {author} {\bibfnamefont {T.}~\bibnamefont {Heckendorf}},
  \bibinfo {author} {\bibfnamefont {S.~B.}\ \bibnamefont {McKagan}}, \bibinfo
  {author} {\bibfnamefont {E.~C.}\ \bibnamefont {Sayre}}, \bibinfo {author}
  {\bibfnamefont {E.~W.}\ \bibnamefont {Schenk}}, \bibinfo {author}
  {\bibfnamefont {C.}~\bibnamefont {Shepherd}},\ and\ \bibinfo {author}
  {\bibfnamefont {L.}~\bibnamefont {Sorell}},\ }\bibfield  {title} {\bibinfo
  {title} {Secondary analysis of teaching methods in introductory physics: {A}
  50 k-student study},\ }\href {https://doi.org/10.1119/1.4964354} {\bibfield
  {journal} {\bibinfo  {journal} {American Journal of Physics}\ }\textbf
  {\bibinfo {volume} {84}},\ \bibinfo {pages} {969} (\bibinfo {year}
  {2016})}\BibitemShut {NoStop}%
\bibitem [{\citenamefont {Kramer}\ \emph {et~al.}(2023)\citenamefont {Kramer},
  \citenamefont {Fuller}, \citenamefont {Watson}, \citenamefont {Castillo},
  \citenamefont {Oliva},\ and\ \citenamefont
  {Potvin}}]{kramer_establishing_2023}%
  \BibitemOpen
  \bibfield  {author} {\bibinfo {author} {\bibfnamefont {L.}~\bibnamefont
  {Kramer}}, \bibinfo {author} {\bibfnamefont {E.}~\bibnamefont {Fuller}},
  \bibinfo {author} {\bibfnamefont {C.}~\bibnamefont {Watson}}, \bibinfo
  {author} {\bibfnamefont {A.}~\bibnamefont {Castillo}}, \bibinfo {author}
  {\bibfnamefont {P.~D.}\ \bibnamefont {Oliva}},\ and\ \bibinfo {author}
  {\bibfnamefont {G.}~\bibnamefont {Potvin}},\ }\bibfield  {title} {\bibinfo
  {title} {Establishing a new standard of care for calculus using trials with
  randomized student allocation},\ }\href
  {https://doi.org/10.1126/science.ade9803} {\bibfield  {journal} {\bibinfo
  {journal} {Science}\ }\textbf {\bibinfo {volume} {381}},\ \bibinfo {pages}
  {995} (\bibinfo {year} {2023})},\ \bibinfo {note} {publisher: American
  Association for the Advancement of Science}\BibitemShut {NoStop}%
\bibitem [{\citenamefont {Freeman}\ \emph {et~al.}(2014)\citenamefont
  {Freeman}, \citenamefont {Eddy}, \citenamefont {McDonough}, \citenamefont
  {Smith}, \citenamefont {Okoroafor}, \citenamefont {Jordt},\ and\
  \citenamefont {Wenderoth}}]{Freeman2014b}%
  \BibitemOpen
  \bibfield  {author} {\bibinfo {author} {\bibfnamefont {S.}~\bibnamefont
  {Freeman}}, \bibinfo {author} {\bibfnamefont {S.~L.}\ \bibnamefont {Eddy}},
  \bibinfo {author} {\bibfnamefont {M.}~\bibnamefont {McDonough}}, \bibinfo
  {author} {\bibfnamefont {M.~K.}\ \bibnamefont {Smith}}, \bibinfo {author}
  {\bibfnamefont {N.}~\bibnamefont {Okoroafor}}, \bibinfo {author}
  {\bibfnamefont {H.}~\bibnamefont {Jordt}},\ and\ \bibinfo {author}
  {\bibfnamefont {M.~P.}\ \bibnamefont {Wenderoth}},\ }\bibfield  {title}
  {\bibinfo {title} {Active learning increases student performance in science,
  engineering, and mathematics},\ }\href
  {https://doi.org/10.1073/pnas.1319030111} {\bibfield  {journal} {\bibinfo
  {journal} {Proceedings of the National Academy of Sciences}\ }\textbf
  {\bibinfo {volume} {111}},\ \bibinfo {pages} {8410} (\bibinfo {year}
  {2014})}\BibitemShut {NoStop}%
\bibitem [{\citenamefont {Meltzer}\ and\ \citenamefont
  {Thornton}(2012)}]{meltzer_resource_2012}%
  \BibitemOpen
  \bibfield  {author} {\bibinfo {author} {\bibfnamefont {D.~E.}\ \bibnamefont
  {Meltzer}}\ and\ \bibinfo {author} {\bibfnamefont {R.~K.}\ \bibnamefont
  {Thornton}},\ }\bibfield  {title} {\bibinfo {title} {Resource {Letter}
  {ALIP}–1: {Active}-{Learning} {Instruction} in {Physics}},\ }\href
  {https://doi.org/10.1119/1.3678299} {\bibfield  {journal} {\bibinfo
  {journal} {American Journal of Physics}\ }\textbf {\bibinfo {volume} {80}},\
  \bibinfo {pages} {478} (\bibinfo {year} {2012})}\BibitemShut {NoStop}%
\bibitem [{\citenamefont {Turpen}\ and\ \citenamefont
  {Finkelstein}(2010)}]{Turpen2010Construction}%
  \BibitemOpen
  \bibfield  {author} {\bibinfo {author} {\bibfnamefont {C.}~\bibnamefont
  {Turpen}}\ and\ \bibinfo {author} {\bibfnamefont {N.~D.}\ \bibnamefont
  {Finkelstein}},\ }\bibfield  {title} {\bibinfo {title} {{The construction of
  different classroom norms during Peer Instruction: Students perceive
  differences}},\ }\href {https://doi.org/10.1103/physrevstper.6.020123}
  {\bibfield  {journal} {\bibinfo  {journal} {Physical Review Special Topics -
  Physics Education Research}\ }\textbf {\bibinfo {volume} {6}},\ \bibinfo
  {pages} {020123+} (\bibinfo {year} {2010})}\BibitemShut {NoStop}%
\bibitem [{\citenamefont {Mazur}(1997)}]{mazur:PI}%
  \BibitemOpen
  \bibfield  {author} {\bibinfo {author} {\bibfnamefont {E.}~\bibnamefont
  {Mazur}},\ }\href@noop {} {\emph {\bibinfo {title} {Peer Instruction: A
  User's Manual}}}\ (\bibinfo  {publisher} {Prentice Hall},\ \bibinfo {year}
  {1997})\BibitemShut {NoStop}%
\bibitem [{\citenamefont {Wilcox}\ \emph {et~al.}(2015)\citenamefont {Wilcox},
  \citenamefont {Kasprzyk},\ and\ \citenamefont {Chini}}]{wilcox}%
  \BibitemOpen
  \bibfield  {author} {\bibinfo {author} {\bibfnamefont {M.}~\bibnamefont
  {Wilcox}}, \bibinfo {author} {\bibfnamefont {C.~C.}\ \bibnamefont
  {Kasprzyk}},\ and\ \bibinfo {author} {\bibfnamefont {J.~J.}\ \bibnamefont
  {Chini}},\ }\bibfield  {title} {\bibinfo {title} {Observing teaching
  assistant differences in tutorials and inquiry-based labs},\ }in\ \href
  {https://doi.org/10.1119/perc.2015.pr.088} {\emph {\bibinfo {booktitle}
  {Physics Education Research Conference 2015}}},\ \bibinfo {series and number}
  {PER Conference}\ (\bibinfo {address} {College Park, MD},\ \bibinfo {year}
  {2015})\ pp.\ \bibinfo {pages} {371--374}\BibitemShut {NoStop}%
\bibitem [{\citenamefont {Potter}\ \emph {et~al.}(2013)\citenamefont {Potter},
  \citenamefont {Webb}, \citenamefont {West}, \citenamefont {Paul},
  \citenamefont {Bowen}, \citenamefont {Weiss}, \citenamefont {Coleman},\ and\
  \citenamefont {De~Leone}}]{Potter:Sixteen}%
  \BibitemOpen
  \bibfield  {author} {\bibinfo {author} {\bibfnamefont {W.~H.}\ \bibnamefont
  {Potter}}, \bibinfo {author} {\bibfnamefont {D.}~\bibnamefont {Webb}},
  \bibinfo {author} {\bibfnamefont {E.~A.}\ \bibnamefont {West}}, \bibinfo
  {author} {\bibfnamefont {C.~A.}\ \bibnamefont {Paul}}, \bibinfo {author}
  {\bibfnamefont {M.}~\bibnamefont {Bowen}}, \bibinfo {author} {\bibfnamefont
  {B.}~\bibnamefont {Weiss}}, \bibinfo {author} {\bibfnamefont
  {L.}~\bibnamefont {Coleman}},\ and\ \bibinfo {author} {\bibfnamefont
  {C.}~\bibnamefont {De~Leone}},\ }\href {http://arxiv.org/abs/1205.6970}
  {\bibinfo {title} {Sixteen years of collaborative learning through active
  sense-making in physics ({CLASP}) at {UC Davis}}} (\bibinfo {year} {2013}),\
  \Eprint {https://arxiv.org/abs/1205.6970} {arXiv:1205.6970} \BibitemShut
  {NoStop}%
\bibitem [{\citenamefont {Scherr}\ \emph {et~al.}(2004)\citenamefont {Scherr},
  \citenamefont {Elby},\ and\ \citenamefont {Goertzen}}]{MarylandTutorials}%
  \BibitemOpen
  \bibfield  {author} {\bibinfo {author} {\bibfnamefont {R.}~\bibnamefont
  {Scherr}}, \bibinfo {author} {\bibfnamefont {A.}~\bibnamefont {Elby}},\ and\
  \bibinfo {author} {\bibfnamefont {R.}~\bibnamefont {Goertzen}},\ }\href@noop
  {} {\emph {\bibinfo {title} {Open Source Tutorials in Physics with Teaching
  Assistant Development}}},\ Vol.\ \bibinfo {volume} {2025}\ (\bibinfo {year}
  {2004})\BibitemShut {NoStop}%
\bibitem [{\citenamefont {Scherr}\ and\ \citenamefont
  {Elby}(2006)}]{MarylandTutorials2}%
  \BibitemOpen
  \bibfield  {author} {\bibinfo {author} {\bibfnamefont {R.}~\bibnamefont
  {Scherr}}\ and\ \bibinfo {author} {\bibfnamefont {A.}~\bibnamefont {Elby}},\
  }\bibfield  {title} {\bibinfo {title} {Enabling informed adaptation of
  reformed instructional materials},\ }in\ \href@noop {} {\emph {\bibinfo
  {booktitle} {Physics Education Research Conference 2006}}},\ \bibinfo
  {series} {PER Conference Invited Paper}, Vol.\ \bibinfo {volume} {883}\
  (\bibinfo {address} {Syracuse, New York},\ \bibinfo {year} {2006})\ pp.\
  \bibinfo {pages} {46--49}\BibitemShut {NoStop}%
\bibitem [{\citenamefont {Paul}\ and\ \citenamefont {West}(2018)}]{Paul2018b}%
  \BibitemOpen
  \bibfield  {author} {\bibinfo {author} {\bibfnamefont {C.}~\bibnamefont
  {Paul}}\ and\ \bibinfo {author} {\bibfnamefont {E.}~\bibnamefont {West}},\
  }\bibfield  {title} {\bibinfo {title} {Using the {Real}-time {Instructor}
  {Observing} {Tool} ({RIOT}) for {Reflection} on {Teaching} {Practice}},\
  }\bibfield  {journal} {\bibinfo  {journal} {Physics Teacher}\ }\textbf
  {\bibinfo {volume} {56}},\ \href {https://doi.org/10.1119/1.5025286}
  {10.1119/1.5025286} (\bibinfo {year} {2018})\BibitemShut {NoStop}%
\bibitem [{\citenamefont {Dancy}\ \emph {et~al.}(2016)\citenamefont {Dancy},
  \citenamefont {Henderson},\ and\ \citenamefont {Turpen}}]{Dancy2016}%
  \BibitemOpen
  \bibfield  {author} {\bibinfo {author} {\bibfnamefont {M.}~\bibnamefont
  {Dancy}}, \bibinfo {author} {\bibfnamefont {C.}~\bibnamefont {Henderson}},\
  and\ \bibinfo {author} {\bibfnamefont {C.}~\bibnamefont {Turpen}},\
  }\bibfield  {title} {\bibinfo {title} {How faculty learn about and implement
  research-based instructional strategies: The case of peer instruction},\
  }\href {https://doi.org/10.1103/physrevphyseducres.12.010110} {\bibfield
  {journal} {\bibinfo  {journal} {Phys. Rev. Phys. Educ. Res.}\ }\textbf
  {\bibinfo {volume} {12}},\ \bibinfo {pages} {010110+} (\bibinfo {year}
  {2016})}\BibitemShut {NoStop}%
\bibitem [{\citenamefont {Smith}\ \emph {et~al.}(2014)\citenamefont {Smith},
  \citenamefont {Vinson}, \citenamefont {Smith}, \citenamefont {Lewin},\ and\
  \citenamefont {Stetzer}}]{smith_campus-wide_2014}%
  \BibitemOpen
  \bibfield  {author} {\bibinfo {author} {\bibfnamefont {M.~K.}\ \bibnamefont
  {Smith}}, \bibinfo {author} {\bibfnamefont {E.~L.}\ \bibnamefont {Vinson}},
  \bibinfo {author} {\bibfnamefont {J.~A.}\ \bibnamefont {Smith}}, \bibinfo
  {author} {\bibfnamefont {J.~D.}\ \bibnamefont {Lewin}},\ and\ \bibinfo
  {author} {\bibfnamefont {M.~R.}\ \bibnamefont {Stetzer}},\ }\bibfield
  {title} {\bibinfo {title} {A {Campus}-{Wide} {Study} of {STEM} {Courses}:
  {New} {Perspectives} on {Teaching} {Practices} and {Perceptions}},\ }\href
  {https://doi.org/10.1187/cbe.14-06-0108} {\bibfield  {journal} {\bibinfo
  {journal} {CBE—Life Sciences Education}\ }\textbf {\bibinfo {volume}
  {13}},\ \bibinfo {pages} {624} (\bibinfo {year} {2014})}\BibitemShut
  {NoStop}%
\bibitem [{\citenamefont {Dancy}\ \emph {et~al.}(2024)\citenamefont {Dancy},
  \citenamefont {Henderson}, \citenamefont {Apkarian}, \citenamefont {Johnson},
  \citenamefont {Stains}, \citenamefont {Raker},\ and\ \citenamefont
  {Lau}}]{dancy_physics_2024}%
  \BibitemOpen
  \bibfield  {author} {\bibinfo {author} {\bibfnamefont {M.}~\bibnamefont
  {Dancy}}, \bibinfo {author} {\bibfnamefont {C.}~\bibnamefont {Henderson}},
  \bibinfo {author} {\bibfnamefont {N.}~\bibnamefont {Apkarian}}, \bibinfo
  {author} {\bibfnamefont {E.}~\bibnamefont {Johnson}}, \bibinfo {author}
  {\bibfnamefont {M.}~\bibnamefont {Stains}}, \bibinfo {author} {\bibfnamefont
  {J.~R.}\ \bibnamefont {Raker}},\ and\ \bibinfo {author} {\bibfnamefont
  {A.}~\bibnamefont {Lau}},\ }\bibfield  {title} {\bibinfo {title} {Physics
  instructors’ knowledge and use of active learning has increased over the
  last decade but most still lecture too much},\ }\href
  {https://doi.org/10.1103/PhysRevPhysEducRes.20.010119} {\bibfield  {journal}
  {\bibinfo  {journal} {Physical Review Physics Education Research}\ }\textbf
  {\bibinfo {volume} {20}},\ \bibinfo {pages} {010119} (\bibinfo {year}
  {2024})}\BibitemShut {NoStop}%
\bibitem [{\citenamefont {Docktor}\ and\ \citenamefont
  {Mestre}(2014)}]{Docktor2014}%
  \BibitemOpen
  \bibfield  {author} {\bibinfo {author} {\bibfnamefont {J.~L.}\ \bibnamefont
  {Docktor}}\ and\ \bibinfo {author} {\bibfnamefont {J.~P.}\ \bibnamefont
  {Mestre}},\ }\bibfield  {title} {\bibinfo {title} {Synthesis of
  discipline-based education research in physics},\ }\href
  {https://doi.org/10.1103/PhysRevSTPER.10.020119} {\bibfield  {journal}
  {\bibinfo  {journal} {Physical Review Special Topics - Physics Education
  Research}\ }\textbf {\bibinfo {volume} {10}},\ \bibinfo {pages} {1} (\bibinfo
  {year} {2014})}\BibitemShut {NoStop}%
\bibitem [{\citenamefont {Saitta}\ \emph {et~al.}(2020)\citenamefont {Saitta},
  \citenamefont {Wilcox}, \citenamefont {James},\ and\ \citenamefont
  {Chini}}]{saitta_views_2020}%
  \BibitemOpen
  \bibfield  {author} {\bibinfo {author} {\bibfnamefont {E.~K.~H.}\
  \bibnamefont {Saitta}}, \bibinfo {author} {\bibfnamefont {M.}~\bibnamefont
  {Wilcox}}, \bibinfo {author} {\bibfnamefont {W.~D.}\ \bibnamefont {James}},\
  and\ \bibinfo {author} {\bibfnamefont {J.~J.}\ \bibnamefont {Chini}},\
  }\bibfield  {title} {\bibinfo {title} {The {Views} of {GTAs} {Impacted} by
  {Cross}-{Tiered} {Professional} {Development}: {Messages} {Intended} and
  {Received}},\ }\href {https://doi.org/10.1007/s40753-020-00115-8} {\bibfield
  {journal} {\bibinfo  {journal} {International Journal of Research in
  Undergraduate Mathematics Education}\ }\textbf {\bibinfo {volume} {6}},\
  \bibinfo {pages} {421} (\bibinfo {year} {2020})}\BibitemShut {NoStop}%
\bibitem [{\citenamefont {Wu}\ \emph {et~al.}(2022)\citenamefont {Wu},
  \citenamefont {Heim}, \citenamefont {Sundstrom}, \citenamefont {Walsh},\ and\
  \citenamefont {Holmes}}]{wu_instructor_2022}%
  \BibitemOpen
  \bibfield  {author} {\bibinfo {author} {\bibfnamefont {D.~G.}\ \bibnamefont
  {Wu}}, \bibinfo {author} {\bibfnamefont {A.~B.}\ \bibnamefont {Heim}},
  \bibinfo {author} {\bibfnamefont {M.}~\bibnamefont {Sundstrom}}, \bibinfo
  {author} {\bibfnamefont {C.}~\bibnamefont {Walsh}},\ and\ \bibinfo {author}
  {\bibfnamefont {N.}~\bibnamefont {Holmes}},\ }\bibfield  {title} {\bibinfo
  {title} {Instructor interactions in traditional and nontraditional labs},\
  }\href {https://doi.org/10.1103/PhysRevPhysEducRes.18.010121} {\bibfield
  {journal} {\bibinfo  {journal} {Physical Review Physics Education Research}\
  }\textbf {\bibinfo {volume} {18}},\ \bibinfo {pages} {010121} (\bibinfo
  {year} {2022})}\BibitemShut {NoStop}%
\bibitem [{\citenamefont {Wan}\ \emph {et~al.}(2020)\citenamefont {Wan},
  \citenamefont {Geraets}, \citenamefont {Doty}, \citenamefont {Saitta},\ and\
  \citenamefont {Chini}}]{wan_characterizing_2020}%
  \BibitemOpen
  \bibfield  {author} {\bibinfo {author} {\bibfnamefont {T.}~\bibnamefont
  {Wan}}, \bibinfo {author} {\bibfnamefont {A.~A.}\ \bibnamefont {Geraets}},
  \bibinfo {author} {\bibfnamefont {C.~M.}\ \bibnamefont {Doty}}, \bibinfo
  {author} {\bibfnamefont {E.~K.~H.}\ \bibnamefont {Saitta}},\ and\ \bibinfo
  {author} {\bibfnamefont {J.~J.}\ \bibnamefont {Chini}},\ }\bibfield  {title}
  {\bibinfo {title} {Characterizing science graduate teaching assistants’
  instructional practices in reformed laboratories and tutorials},\ }\href
  {https://doi.org/10.1186/s40594-020-00229-0} {\bibfield  {journal} {\bibinfo
  {journal} {International Journal of STEM Education}\ }\textbf {\bibinfo
  {volume} {7}},\ \bibinfo {pages} {30} (\bibinfo {year} {2020})}\BibitemShut
  {NoStop}%
\bibitem [{\citenamefont {Stang}\ and\ \citenamefont
  {Roll}(2014)}]{Stang2014Interactions}%
  \BibitemOpen
  \bibfield  {author} {\bibinfo {author} {\bibfnamefont {J.~B.}\ \bibnamefont
  {Stang}}\ and\ \bibinfo {author} {\bibfnamefont {I.}~\bibnamefont {Roll}},\
  }\bibfield  {title} {\bibinfo {title} {{Interactions between teaching
  assistants and students boost engagement in physics labs}},\ }\bibfield
  {journal} {\bibinfo  {journal} {Physical Review Special Topics - Physics
  Education Research}\ }\textbf {\bibinfo {volume} {10}},\ \href
  {https://doi.org/10.1103/physrevstper.10.020117}
  {10.1103/physrevstper.10.020117} (\bibinfo {year} {2014})\BibitemShut
  {NoStop}%
\bibitem [{\citenamefont {Sawada}\ \emph {et~al.}(2002)\citenamefont {Sawada},
  \citenamefont {Piburn}, \citenamefont {Judson}, \citenamefont {Turley},
  \citenamefont {Falconer}, \citenamefont {Benford},\ and\ \citenamefont
  {Bloom}}]{rtop}%
  \BibitemOpen
  \bibfield  {author} {\bibinfo {author} {\bibfnamefont {D.}~\bibnamefont
  {Sawada}}, \bibinfo {author} {\bibfnamefont {M.~D.}\ \bibnamefont {Piburn}},
  \bibinfo {author} {\bibfnamefont {E.}~\bibnamefont {Judson}}, \bibinfo
  {author} {\bibfnamefont {J.}~\bibnamefont {Turley}}, \bibinfo {author}
  {\bibfnamefont {K.}~\bibnamefont {Falconer}}, \bibinfo {author}
  {\bibfnamefont {R.}~\bibnamefont {Benford}},\ and\ \bibinfo {author}
  {\bibfnamefont {I.}~\bibnamefont {Bloom}},\ }\bibfield  {title} {\bibinfo
  {title} {{Measuring Reform Practices in Science and Mathematics Classrooms:
  The Reformed Teaching Observation Protocol}},\ }\href
  {https://doi.org/10.1111/j.1949-8594.2002.tb17883.x} {\bibfield  {journal}
  {\bibinfo  {journal} {School Science and Mathematics}\ }\textbf {\bibinfo
  {volume} {102}},\ \bibinfo {pages} {245} (\bibinfo {year}
  {2002})}\BibitemShut {NoStop}%
\bibitem [{\citenamefont {Smith}\ \emph {et~al.}(2013)\citenamefont {Smith},
  \citenamefont {Jones}, \citenamefont {Gilbert},\ and\ \citenamefont
  {Wieman}}]{copus}%
  \BibitemOpen
  \bibfield  {author} {\bibinfo {author} {\bibfnamefont {M.~K.}\ \bibnamefont
  {Smith}}, \bibinfo {author} {\bibfnamefont {F.~H.~M.}\ \bibnamefont {Jones}},
  \bibinfo {author} {\bibfnamefont {S.~L.}\ \bibnamefont {Gilbert}},\ and\
  \bibinfo {author} {\bibfnamefont {C.~E.}\ \bibnamefont {Wieman}},\ }\bibfield
   {title} {\bibinfo {title} {{The Classroom Observation Protocol for
  Undergraduate STEM (COPUS): A New Instrument to Characterize University STEM
  Classroom Practices}},\ }\href {https://doi.org/10.1187/cbe.13-08-0154}
  {\bibfield  {journal} {\bibinfo  {journal} {CBE-Life Sciences Education}\
  }\textbf {\bibinfo {volume} {12}},\ \bibinfo {pages} {618} (\bibinfo {year}
  {2013})}\BibitemShut {NoStop}%
\bibitem [{\citenamefont {Hora}\ \emph {et~al.}(2013)\citenamefont {Hora},
  \citenamefont {Oleson},\ and\ \citenamefont {Ferrare}}]{tdop}%
  \BibitemOpen
  \bibfield  {author} {\bibinfo {author} {\bibfnamefont {M.~T.}\ \bibnamefont
  {Hora}}, \bibinfo {author} {\bibfnamefont {A.}~\bibnamefont {Oleson}},\ and\
  \bibinfo {author} {\bibfnamefont {J.~J.}\ \bibnamefont {Ferrare}},\
  }\bibfield  {title} {\bibinfo {title} {{Teaching Dimensions Observation
  Protocol (TDOP) User's Manual}},\ }\href
  {http://tdop.wceruw.org/Document/TDOP-Users-Guide.pdf} {\bibfield  {journal}
  {\bibinfo  {journal} {Madison: Wisconsin Center for Education Research}\ }
  (\bibinfo {year} {2013})}\BibitemShut {NoStop}%
\bibitem [{\citenamefont {Velasco}\ \emph {et~al.}(2016)\citenamefont
  {Velasco}, \citenamefont {Knedeisen}, \citenamefont {Xue}, \citenamefont
  {Vickrey}, \citenamefont {Abebe},\ and\ \citenamefont {Stains}}]{lopus}%
  \BibitemOpen
  \bibfield  {author} {\bibinfo {author} {\bibfnamefont {J.~B.}\ \bibnamefont
  {Velasco}}, \bibinfo {author} {\bibfnamefont {A.}~\bibnamefont {Knedeisen}},
  \bibinfo {author} {\bibfnamefont {D.}~\bibnamefont {Xue}}, \bibinfo {author}
  {\bibfnamefont {T.~L.}\ \bibnamefont {Vickrey}}, \bibinfo {author}
  {\bibfnamefont {M.}~\bibnamefont {Abebe}},\ and\ \bibinfo {author}
  {\bibfnamefont {M.}~\bibnamefont {Stains}},\ }\bibfield  {title} {\bibinfo
  {title} {Characterizing instructional practices in the laboratory: The
  laboratory observation protocol for undergraduate stem},\ }\href
  {https://doi.org/10.1021/acs.jchemed.6b00062} {\bibfield  {journal} {\bibinfo
   {journal} {Journal of Chemical Education}\ }\textbf {\bibinfo {volume}
  {93}},\ \bibinfo {pages} {1191} (\bibinfo {year} {2016})},\ \Eprint
  {https://arxiv.org/abs/https://doi.org/10.1021/acs.jchemed.6b00062}
  {https://doi.org/10.1021/acs.jchemed.6b00062} \BibitemShut {NoStop}%
\bibitem [{\citenamefont {Madsen}\ \emph {et~al.}(2019)\citenamefont {Madsen},
  \citenamefont {McKagan}, \citenamefont {Sayre},\ and\ \citenamefont
  {Paul}}]{Madsen2019}%
  \BibitemOpen
  \bibfield  {author} {\bibinfo {author} {\bibfnamefont {A.}~\bibnamefont
  {Madsen}}, \bibinfo {author} {\bibfnamefont {S.~B.}\ \bibnamefont {McKagan}},
  \bibinfo {author} {\bibfnamefont {E.~C.}\ \bibnamefont {Sayre}},\ and\
  \bibinfo {author} {\bibfnamefont {C.~A.}\ \bibnamefont {Paul}},\ }\bibfield
  {title} {\bibinfo {title} {Resource {Letter} {RBAI}-2: {Research}-based
  assessment instruments: {Beyond} physics topics},\ }\bibfield  {journal}
  {\bibinfo  {journal} {American Journal of Physics}\ }\href
  {https://doi.org/https://doi.org/10.1119/1.5094139}
  {https://doi.org/10.1119/1.5094139} (\bibinfo {year} {2019})\BibitemShut
  {NoStop}%
\bibitem [{\citenamefont {Cohen}(1960)}]{cohen_1960_coefficient}%
  \BibitemOpen
  \bibfield  {author} {\bibinfo {author} {\bibfnamefont {J.}~\bibnamefont
  {Cohen}},\ }\bibfield  {title} {\bibinfo {title} {{A Coefficient of Agreement
  for Nominal Scales}},\ }\href {https://doi.org/10.1177/001316446002000104}
  {\bibfield  {journal} {\bibinfo  {journal} {Educational and Psychological
  Measurement}\ }\textbf {\bibinfo {volume} {20}},\ \bibinfo {pages} {37}
  (\bibinfo {year} {1960})}\BibitemShut {NoStop}%
\bibitem [{\citenamefont {J.~Richard~Landis}(1977)}]{landisandkoch}%
  \BibitemOpen
  \bibfield  {author} {\bibinfo {author} {\bibfnamefont {G.~G.~K.}\
  \bibnamefont {J.~Richard~Landis}},\ }\bibfield  {title} {\bibinfo {title}
  {The measurement of observer agreement for categorical data},\ }\href
  {http://www.jstor.org/stable/2529310} {\bibfield  {journal} {\bibinfo
  {journal} {Biometrics}\ }\textbf {\bibinfo {volume} {33}},\ \bibinfo {pages}
  {159} (\bibinfo {year} {1977})}\BibitemShut {NoStop}%
\bibitem [{\citenamefont {Von~Korff}\ \emph {et~al.}(2015)\citenamefont
  {Von~Korff}, \citenamefont {Zhan}, \citenamefont {Vaishnav}, \citenamefont
  {Chini}, \citenamefont {Warneke},\ and\ \citenamefont {Sengul}}]{VonKorff}%
  \BibitemOpen
  \bibfield  {author} {\bibinfo {author} {\bibfnamefont {J.~S.}\ \bibnamefont
  {Von~Korff}}, \bibinfo {author} {\bibfnamefont {C.}~\bibnamefont {Zhan}},
  \bibinfo {author} {\bibfnamefont {b.}~\bibnamefont {Vaishnav}}, \bibinfo
  {author} {\bibfnamefont {J.~J.}\ \bibnamefont {Chini}}, \bibinfo {author}
  {\bibfnamefont {A.}~\bibnamefont {Warneke}},\ and\ \bibinfo {author}
  {\bibfnamefont {O.}~\bibnamefont {Sengul}},\ }\bibfield  {title} {\bibinfo
  {title} {he use of representations in evidence-based and non-evidence-based
  physics activities},\ }in\ \href {https://doi.org/10.1119/perc.2015.pr.083}
  {\emph {\bibinfo {booktitle} {Physics Education Research Conference 2015}}},\
  \bibinfo {series and number} {PER Conference}\ (\bibinfo {address} {College
  Park, MD},\ \bibinfo {year} {2015})\ pp.\ \bibinfo {pages}
  {351--354}\BibitemShut {NoStop}%
\bibitem [{\citenamefont {Fleiss}(1981)}]{fleiss}%
  \BibitemOpen
  \bibfield  {author} {\bibinfo {author} {\bibfnamefont {J.}~\bibnamefont
  {Fleiss}},\ }\href@noop {} {\emph {\bibinfo {title} {Statistical methods for
  rates and proportions}}},\ \bibinfo {edition} {2nd}\ ed.\ (\bibinfo
  {publisher} {Wiley},\ \bibinfo {address} {New York},\ \bibinfo {year}
  {1981})\BibitemShut {NoStop}%
\bibitem [{\citenamefont {Wilcox}\ \emph {et~al.}(2016)\citenamefont {Wilcox},
  \citenamefont {Yang},\ and\ \citenamefont {Chini}}]{wilcox_quicker_2016}%
  \BibitemOpen
  \bibfield  {author} {\bibinfo {author} {\bibfnamefont {M.}~\bibnamefont
  {Wilcox}}, \bibinfo {author} {\bibfnamefont {Y.}~\bibnamefont {Yang}},\ and\
  \bibinfo {author} {\bibfnamefont {J.~J.}\ \bibnamefont {Chini}},\ }\bibfield
  {title} {\bibinfo {title} {Quicker method for assessing influences on
  teaching assistant buy-in and practices in reformed courses},\ }\href
  {https://doi.org/10.1103/PhysRevPhysEducRes.12.020123} {\bibfield  {journal}
  {\bibinfo  {journal} {Physical Review Physics Education Research}\ }\textbf
  {\bibinfo {volume} {12}},\ \bibinfo {pages} {020123} (\bibinfo {year}
  {2016})},\ \bibinfo {note} {publisher: American Physical Society}\BibitemShut
  {NoStop}%
\bibitem [{\citenamefont {Kim}\ and\ \citenamefont
  {Pak}(2002)}]{Kim2002Students}%
  \BibitemOpen
  \bibfield  {author} {\bibinfo {author} {\bibfnamefont {E.}~\bibnamefont
  {Kim}}\ and\ \bibinfo {author} {\bibfnamefont {S.-J.}\ \bibnamefont {Pak}},\
  }\bibfield  {title} {\bibinfo {title} {Students do not overcome conceptual
  difficulties after solving 1000 traditional problems},\ }\href
  {https://doi.org/10.1119/1.1484151} {\bibfield  {journal} {\bibinfo
  {journal} {American Journal of Physics}\ }\textbf {\bibinfo {volume} {70}},\
  \bibinfo {pages} {759} (\bibinfo {year} {2002})},\ \bibinfo {note}
  {publisher: AAPT}\BibitemShut {NoStop}%
\end{thebibliography}%

\end{document}